\documentclass[aps,pra,twocolumn,superscriptaddress,longbibliography]{revtex4-2}
\usepackage{graphicx}
\usepackage{natbib}
\usepackage{amsmath}
\usepackage{amsfonts}

\newcommand{\ket}[1]{\ensuremath{\left|{#1}\right\rangle}}
\newcommand{\bra}[1]{\ensuremath{\left\langle{#1}\right|}}
\newcommand{\figfolder}[1]{}

\usepackage[usenames,dvipsnames]{xcolor}
\usepackage{hyperref}
\hypersetup{
	pdfnewwindow=true, 
	colorlinks=true, 
	linkcolor=MidnightBlue,
	citecolor=Magenta,
	urlcolor=RoyalBlue
}
\usepackage{changes}

\let\sss=\scriptscriptstyle

\newcommand{\addrA}{Beijing Key Laboratory of Fault-Tolerant Quantum Computing, Beijing Academy of Quantum Information Sciences, Beijing 100193, China} 
\newcommand{\addrB}{Beijing National Laboratory for Condensed Matter Physics, Institute of Physics, Chinese Academy of Sciences, Beijing 100190, China} 
\newcommand{\addrC}{School of Physical Sciences, University of Chinese Academy of Sciences, Beijing 100049, China} 
\newcommand{\addrD}{Hefei National Laboratory, Hefei 230088, China}

\begin{document}


\title{Microwave-activated high-fidelity three-qubit gate scheme for fixed-frequency superconducting qubits}

\author{Kui Zhao}
	\affiliation{\addrA}
	
\author{Wei-Guo Ma}
	\affiliation{\addrB}
	\affiliation{\addrC}
	
\author{Ziting Wang}
	\affiliation{\addrA}

\author{Hao Li}
	\affiliation{\addrA}
	
\author{Kaixuan Huang}
	\affiliation{\addrA}
	
\author{Yun-Hao Shi}
    \email{yhshi@iphy.ac.cn}
	\affiliation{\addrB}
	
\author{Kai Xu}
    \email{kaixu@iphy.ac.cn}
	\affiliation{\addrB}
	\affiliation{\addrA}
	\affiliation{\addrC}
	\affiliation{\addrD}
    
\author{Heng Fan}
    \email{hfan@iphy.ac.cn}
	\affiliation{\addrB}
	\affiliation{\addrA}
	\affiliation{\addrC}
	\affiliation{\addrD}


\begin{abstract}

Scalable superconducting quantum processors require balancing critical constraints in coherence, control complexity, and spectral crowding. Fixed-frequency architectures suppress flux noise and simplify control via all-microwave operations but remain limited by residual ZZ crosstalk. Here we propose a microwave-activated three-qubit gate protocol for fixed-frequency transmon qubits in the large-detuning regime ($|\Delta| \gg g$), leveraging the third-order nonlinear interaction to coherently exchange $\ket{001} \leftrightarrow \ket{110}$ states. By incorporating a phase-compensated optimization protocol, numerical simulations demonstrate a high average gate fidelity exceeding $99.9\%$.
Systematic error analysis identifies static long-range ZZ coupling as the dominant error source in multi-qubit systems, which can be suppressed via operations in the large-detuning regime ($\sim 1$ GHz). 
The protocol maintains process fidelities exceeding 98\% under decoherence, while demonstrating intrinsic robustness to fabrication-induced parameter variations and compatibility with existing all-microwave two-qubit gate architectures. This hardware-efficient strategy advances scalable quantum computing systems by improving coherence properties, reducing spectral congestion, and expanding the experimental toolkit for error-resilient quantum operations in the noisy intermediate-scale quantum era.

\end{abstract}

\maketitle

\section{Introduction\label{sec:Introduction}}

Achieving scalable quantum computation requires addressing two concurrent demands: mitigating decoherence in noisy systems and implementing complex algorithms with minimal physical resources. Despite~significant progress in quantum error correction~\cite{Krinner2022Nature,PhysRevLett.129.030501,Ni2023Nature,Acharya2025Willow}, current quantum technologies still fall short of realizing fault-tolerant universal quantum computation. This limitation becomes particularly critical under the noisy intermediate-scale quantum (NISQ) paradigm~\cite{Preskill2018quantumcomputingin,RevModPhys.94.015004,Cheng2023}, where full-scale error correction is infeasible, underscoring the need for quantum architectures that support precise two- and multi-qubit operations while maintaining error resilience.

Fixed-frequency superconducting architectures gained prominence as a platform for NISQ devices~\cite{PhysRevB.81.134507,PhysRevLett.107.080502,PhysRevA.93.060302,PhysRevApplied.12.064013,PRXQuantum.2.040336,PhysRevLett.127.130501,PhysRevLett.127.200502,PhysRevLett.129.060501,li_hardware-efficient_2024,PRXQuantum.5.020338,Le2023}, offering enhanced coherence times and simplified control electronics compared to flux-tunable designs~\cite{PhysRevA.90.022307,PhysRevA.102.042605,PhysRevApplied.10.054062,PhysRevLett.125.240502,PhysRevLett.125.240503,PhysRevLett.123.120502,PhysRevLett.126.220502,PhysRevX.11.021058,PhysRevLett.125.240502,PRXQuantum.4.010314,PhysRevApplied.18.034038,PhysRevApplied.16.024037,PhysRevApplied.16.024037,PhysRevX.14.041050,PhysRevApplied.23.024059}. These architectures avoid the decoherence channels associated with frequency tuning while enabling all-microwave gate protocols~\cite{PhysRevA.102.062408,PhysRevA.76.042319,PhysRevA.102.042605}. However, their fixed couplings introduce persistent ZZ crosstalk~\cite{PhysRevLett.127.200502,PhysRevLett.127.130501,PhysRevLett.129.060501}, which degrades gate fidelity and restricts operational bandwidth. Conventional approaches such as cross-resonance gates ~\cite{PhysRevA.93.060302,PhysRevApplied.12.064013,PRXQuantum.2.040336} further exacerbate these issues by requiring qubit detunings smaller than their anharmonicities~\!\cite{PhysRevB.81.134507,PhysRevA.93.060302,PhysRevLett.109.060501},~thereby intensifying spectral crowding and imposing stringent fabrication tolerances~\cite{PhysRevResearch.4.023079}. Emerging large-detuning architectures help resolve these conflicts by operating in dispersive coupling regimes~\cite{PhysRevApplied.14.044039,PhysRevApplied.22.034007}. These systems effectively suppress static ZZ interactions and enable high-fidelity, microwave-driven controlled-phase (CPhase) and controlled-Z (CZ) gates. This capability not only preserves the intrinsic coherence advantages of fixed-frequency architectures but also enhances their scalability.

However, extending these advances to large-scale multi-qubit operations poses a critical challenge. Three-qubit gates, such as the Toffoli and Fredkin gates, play crucial roles in quantum algorithms~\cite{PhysRevA.52.R2493,PhysRevLett.121.010501}, quantum simulations~\cite{RevModPhys.86.153}, and error correction protocols~\cite{PhysRevLett.81.2152,science.1203329,PhysRevA.63.052314,PhysRevLett.111.090505}. Direct hardware-level implementations of such gates, as opposed to decomposed sequences, offer substantial reductions in circuit depth and enhanced operational flexibility~\cite{PhysRevA.88.010304}. Recent advances in superconducting quantum circuits have propelled significant progress in implementing three-qubit gates, such as iToffoli~\cite{iToffoli_high-fidelity_2022,i-Toffoli-APL,Fedorov2012}, CCZ~\cite{reed_realization_2012,Nguyen2024,liu2025directimplementationhighfidelitythreequbit,LiShaowei2019,Nguyen2024dualbosonicladder,PhysRevApplied.19.044001,PhysRevApplied.21.044035}, and iFredkin~\cite{PhysRevA.101.062312,li_hardware-efficient_2024}, by leveraging strategies including simultaneous pairwise interactions~\cite{PRXQuantum.2.040348,warren_extensive_2023,PhysRevApplied.21.034018}, high-level (qutrit, qudit)-assisted protocols~\cite{Fedorov2012,reed_realization_2012,li_hardware-efficient_2024,Nguyen2024} and coupler (cavity)-assisted architectures~\cite{science.1208517,PhysRevB.96.024504,reed_realization_2012}. Despite these advancements, practical implementations remain constrained by suboptimal coupling strengths, prolonged high-energy state occupation, and frequency crowding from spurious interactions, which respectively limit gate speeds, amplify decoherence errors, and introduce fidelity-limiting crosstalk.

In this work, we investigate the implementation of three-qubit gates in large-detuning fixed-frequency architectures. Building on insights from prior studies~\cite{PhysRevLett.130.260601,PhysRevX.10.021038}, we propose a microwave-activated three-qubit gate protocol specifically optimized for large-detuning architectures. The protocol utilizes the intrinsic third-order nonlinearity of transmon qubits to mediate a controlled state exchange ($|001\rangle \leftrightarrow |110\rangle$) using a single microwave drive, enabling selective three-qubit operations. 
The large-detuning gate scheme is inherently compatible with established all-microwave CPhase and CZ gate protocols~\cite{PhysRevApplied.14.044039}, which operate via Raman transition pathways. Furthermore, we demonstrate that all-microwave, continuous-wave Stark drives—employing off-resonant microwave tones—can be used to induce conditional Stark shifts. By adjusting the phase of these drives, we achieve complete suppression of ZZ crosstalk in fixed-transmon architectures, while enabling dynamic ZZ modulation for the implementation of high-fidelity CPhase and CZ gates. This approach ensures seamless integration with standard two-qubit gate frameworks.
By combining this protocol with established CZ gate techniques, we demonstrate efficient synthesis of three-qubit gates, such as the iFredkin gate, while retaining the spectral isolation and scalability intrinsic to fixed-frequency architectures. Our work bridges the gap between robust two-qubit operations and high-fidelity multi-qubit control, advancing the toolkit for scalable quantum computing in the NISQ era.

The paper is organized as follows. Section~\!\ref{sec:sys} introduces the superconducting circuit model and establishes the physical mechanism for microwave-activated state-transfer processes, forming the theoretical foundation of the three-qubit gate protocol. Section~\!\ref{sec:transition} systematically explores the dynamics of microwave-driven transitions, elucidating critical relationships between control parameters and coherent population transfer. Section~\!\ref{sec:Gateperformance} quantifies gate performance through numerical simulations, analyzing average fidelity across parameter variations (coupling strengths, detunings) to identify optimal operational regimes. Conclusions and broader implications are summarized in Section~\!\ref{sec:Conclusion}. Appendices provide supplementary technical analyses. Appendix~\!\ref{app:4WM} discusses the origin of the third-order nonlinear interaction for realizing the three-qubit gate. Appendix~\!\ref{sec:RWA_validity} examines the validity of the rotating wave approximation (RWA) in our simulations. Appendix~\!\ref{app:CWStark} details the implementation and performance of continuous-wave Stark drives for dynamic ZZ coupling management in fixed-frequency transmon architectures.
Appendix~\!\ref{app:EvolutionOperator} details evolution operator simulations, and Appendix~\!\ref{app:ProcessFidelity} addresses process fidelity evaluation methodologies.

\section{system and model\label{sec:sys}}

\begin{figure}[t]
        \centering
	\includegraphics[width = 0.8\linewidth]{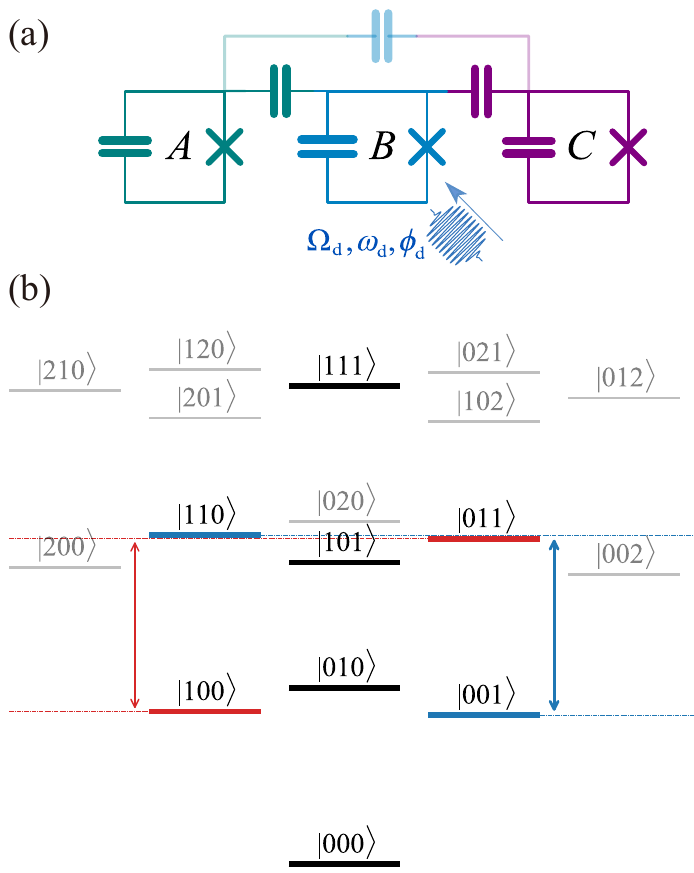}
	\caption{Circuit diagram and microwave-activated scheme. (a) Circuit schematic of three fixed-frequency transmon qubits ($q_{\sss A}$, $q_{\sss B}$, $q_{\sss C}$) with nearest-neighbor capacitive couplings. The microwave-activated transition is driven by an external microwave field applied to $q_{\sss B}$, characterized by amplitude $\Omega_d$ and frequency $\omega_d$ and phase $\phi_{d}$. (b) Energy-level diagram of the coupled system with a maximum total excitation number below four, where $\ket{n_{\sss A} n_{\sss B} n_{\sss C}}$ denotes the collective state with $n_{\sss A,B,C}$ excitations in each qubit. Within the computational subspace spanned by $\{\ket{000}, \ket{001}, \ket{010}, \ket{011}, \ket{100}, \ket{101}, \ket{110}, \ket{111}\}$. The microwave-activated transition mediates resonant state exchange between $\ket{001}\leftrightarrow\ket{110}$ (blue branch) and $\ket{100}\leftrightarrow\ket{011}$ (red branch), as depicted by the corresponding blue and red double arrows in the energy-level diagram.}
    \label{fig:MAS-Circuit}
\end{figure}

\begin{figure}[t]
	\centering
		\includegraphics[width=0.97\linewidth]{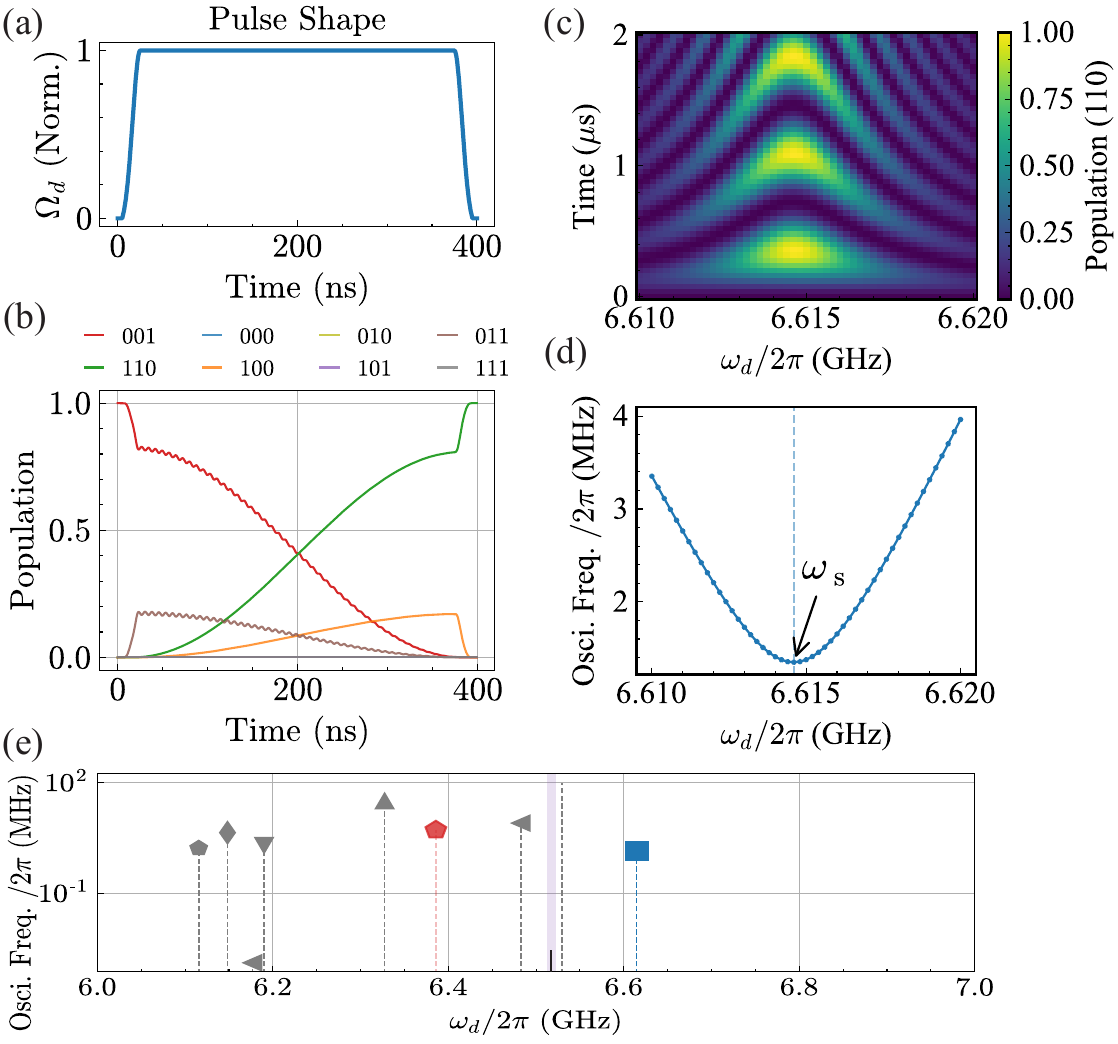}
		\caption{Pulse shape profile and microwave-activated state-transfer dynamics. (a) Numerically simulated pulse envelope (blue trace) featuring a flat-top profile with Gaussian edges ($\sigma=10$ ns, total edge duration $4\sigma$). 
		(b) microwave-activated resonant state exchange between $\ket{001}$ and $\ket{110}$, achieving $>99.9\%$ population transfer fidelity from the initial $\ket{001}$ state.
		Notably, all non-target computational states remain unpopulated ($<0.1\%$), highlighting intrinsic state selectivity of the protocol. (c) Population of the $\ket{110}$ state for the microwave-activated transition as a function of drive frequency $\omega_d$ and duration time, obtained by a numerical simulation of the transmon Hamiltonian (Eq.~\!\eqref{eq:eq003}) when the system is initialized in $\ket{001}$ state. (d) Extracted oscillation frequencies from (c) align with the theoretical value of approximately $1.25$ MHz.
		(e) Broadband spectroscopy. The spectrum reveals spectral features, including a primary blue brand resonance (blue) , a red branch sideband (red) , sidebands around $\omega_{\sss B}$ and distant features.
		}
		\label{fig:pulse}
\end{figure}

The superconducting circuit system comprises three fixed-frequency transmon qubits with adjacent capacitive couplings, as schematically depicted in Fig.~\!\ref{fig:MAS-Circuit}(a), governed by the Hamiltonian (hereafter $\hbar = 1$)
\begin{equation}\label{eq:eq001}
\begin{aligned}
    \hat{H}_0 =& \sum_i \left(\omega_i \hat{a}^\dagger_i \hat{a}_i + \frac{\alpha_i}{2}\hat{a}^\dagger_i\hat{a}^\dagger_i\hat{a}_i\hat{a}_i\right) + \\& \sum_{i,j} g_{ij}(\hat{a}_i+\hat{a}^\dagger_i)(\hat{a}_j+\hat{a}^\dagger_j),   
\end{aligned}
\end{equation}
where $\omega_i$, $\alpha_i$, and $g_{ij}$  denote qubit frequencies, anharmonicities and pairwise coupling strengths, respectively, for $i,j \in \{A,B,C\}$. The $\hat{a}_i$ ($\hat{a}^\dagger_i$) denotes the annihilation (creation) operator for $q_i$. Figure~\!\ref{fig:MAS-Circuit}(b) shows the energy levels of the total system. The inherent third-order nonlinearity of the transmon qubits allows the microwave drive to modulate the system’s Hamiltonian through a four-wave mixing process, facilitating coherent transitions between the distinct states~\!\cite{science.aaa2085two-photon,PhysRevX.10.021038,PhysRevLett.130.260601}. The microwave-activated interaction can be activated by directly driving the central qubit $q_{\sss B}$, i.e., $\hat{H}_\mathrm{d}=\Omega_d(t)\cos{(\omega_d t + \phi_{d})}(\hat{a}^\dagger_{\sss B} + \hat{a}_{\sss B})$,
where $\Omega_d(t)$ denotes the time-dependent pulse envelope, $\omega_d$ represents the drive frequency, and $\phi_{d}$ is the drive microwave phase. Here we set $\Omega_d(t)$ as the flat-top Gaussian waveform (Fig.~\!\ref{fig:pulse}(a)), parametrized by the maximal amplitude $\Omega_d$, the duration $\tau$ and the Gaussian edge $\sigma$ ($\sigma = 10$~ns, total ramp edge duration $4\sigma$). In the rotating frame, the Hamiltonian can be simplified as
\begin{equation}\label{eq:eq003}
\begin{aligned}
        &\hat{H}_{\mathrm{RF}} = \sum_i \left[(\omega_i-\omega_d) \hat{a}^\dagger_i \hat{a}_i + \frac{\alpha_i}{2}\hat{a}^\dagger_i\hat{a}^\dagger_i\hat{a}_i\hat{a}_i\right] \\
        &+ \sum_{i,j} g_{ij}(\hat{a}^\dagger_i \hat{a}_j + \hat{a}_i\hat{a}^\dagger_j) 
        + \frac{1}{2} \left[\Omega_d(t)e^{-i\phi_{d}} \hat{a}^\dagger_{\sss B} + \Omega^{*}_d(t)e^{i\phi_{d}} \hat{a}_{\sss B} \right].
\end{aligned}
\end{equation}
The above rotating-frame Hamiltonian serves as the foundation for our analysis. In the dispersive regime (where $ |g_{\sss A(C)B}/\Delta_{\sss A(C)B}| \ll 1 $), the inherent nonlinearity of the qubit enables four-wave mixing processes that coherently couple three qubits with a single drive photon~\!\cite{science.aaa2085two-photon,PhysRevX.10.021038}. When the frequency of the external microwave drive precisely matches the energy-level difference between the computational states $ \ket{001} \leftrightarrow \ket{110} $ or $ \ket{100} \leftrightarrow \ket{011} $ (graphically represented by the red and blue double arrows in Fig.~\ref{fig:MAS-Circuit}(b)), the system mediates microwave-activated effective transitions between these states. The perturbation analysis under the rotating-wave approximation yields an oscillation frequency $\nu$ for the $\ket{001} \leftrightarrow \ket{110}$ transition~\!\cite{PhysRevLett.130.260601}
\begin{equation}\label{eq:eq004}
    \begin{aligned}
        \nu &\approx 2\bra{001}\hat{H}_{\mathrm{RF}}\ket{110} = \frac{2g_{\sss AB} g_{\sss BC} \alpha_{\sss B} \Omega_d}{\Delta_{\sss AC}\Delta_{\sss BC}(\Delta_{\sss BA}+\alpha_{\sss B})},
    \end{aligned}
\end{equation}
where $ \Delta_{ij} = \omega_i - \omega_j $. Without loss of generality, we adopt the convention $ \omega_{\sss A} > \omega_{\sss C} $ throughout the following analysis.

Note that the microwave-activated transitions enable controlled state exchange between $\ket{001} \leftrightarrow \ket{110}$ (blue branch) and $\ket{100} \leftrightarrow \ket{011}$ (red branch), distinguished by their unique transition frequencies. The red branch operates at a frequency below $\omega_{\sss B}$, making it susceptible to off-resonant effects such as
single-photon ($\ket{101}\leftrightarrow\ket{210}$) and two-photon ($\ket{000}\leftrightarrow\ket{020}$, $\ket{001}\leftrightarrow\ket{120}$) transitions~\!\cite{PhysRevLett.130.260601}.
In contrast, the blue branch operates at a frequency above $\omega_{\sss B}$, which suppresses spurious transitions on the high-frequency side of $\omega_{\sss B}$ due to the negative anharmonicity of the transmon. This intrinsic spectral isolation establishes the blue branch as the superior platform for implementing high-fidelity gates. 
Based on the microwave-activated transitions, which were pioneered by Shirai et al.~\!\cite{PhysRevLett.130.260601} in their realization of a two-qubit CZ gate within a fixed-frequency architecture, our work explores the inherent four-wave mixing mechanism behind it for direct three-qubit gate implementation. By exploiting the natural $\ket{001}~\!\leftrightarrow~\!\ket{110}$ state exchange, we demonstrate that this physical process can be natively extended to implement high-fidelity multi-qubit operations.
This approach retains the inherent advantages of fixed-frequency architectures, such as minimized frequency crowding and robust coherence, while expanding their functional versatility for multi-qubit operations.
Our work highlights the practical applicability and technical feasibility of leveraging intrinsic multi-qubit interactions, offering a promising pathway toward enhanced scalability and operational flexibility in superconducting quantum processors.

\section{microwave-activated transition\label{sec:transition}}

In this section, we investigate the dynamics of microwave-activated transitions in a three-transmon circuit through numerical simulations. The simulations, implemented using QuTiP~\!\cite{JOHANSSON20121760}, model a capacitively coupled fixed-frequency transmon system governed by the Hamiltonian in Eq.~\!\eqref{eq:eq003}.
Each qubit is truncated at the fourth excited state to balance computational efficiency with physical accuracy, capturing essential nonlinear behavior while maintaining numerical stability. The parameters are designed to align with experimental implementations~\!\cite{PhysRevLett.130.260601} while strategically enhancing the detuning between nearest-neighbor (NN) qubits, with the corresponding frequencies $\omega_{\sss A,B,C}/2\pi = \{5.641, 6.517, 5.507\}$ GHz, anharmonicities $\alpha_{\sss A,B,C}/2\pi = \{-300, -381, -303\}$~MHz, NN coupling strengths $g_{\sss AB}/2\pi = 40$~MHz and $g_{\sss BC}/2\pi = 31$~MHz, and a next-nearest-neighbor (NNN) coupling $g_{\sss AC}/2\pi = 1.9$~MHz. The above parameter configuration preserves the intrinsic scalability of fixed-frequency architectures while maintaining compatibility with established high-fidelity CPhase gate protocols~\!\cite{PhysRevApplied.14.044039}.

Despite operating in a large-detuning regime, the fixed capacitive couplings inherently introduce static ZZ interactions, with NN ZZ couplings calculated as $\xi_{\sss AB}/2\pi \approx -3.7$ MHz and $\xi_{\sss BC}/2\pi \approx -1.5$ MHz, in agreement with perturbation formula~\!\cite{PhysRevApplied.10.054062} 
\begin{equation}\label{eq:eqZZNNij} 
    \xi_{ij} = \frac{2g_{ij}^2(\alpha_i+\alpha_j)}{(\Delta_{ij}+\alpha_i)(\Delta_{ij}-\alpha_j)}.
\end{equation}
Note that the large detuning ($\sim 1.0$ GHz) between NN qubits yields $\xi_{\sss AC}/2\pi \approx 6.9$ kHz, which implies that the long-range ZZ interaction can be significant suppressed. 
Although static ZZ couplings may introduce parasitic CPhase phase errors (see subsequent numerical analyses), recent advances in fixed-coupling architectures demonstrate that simultaneous off-resonant continuous-wave (CW) Stark tones can further suppress these residual interactions~\!\cite{PhysRevLett.127.200502,PhysRevLett.129.060501,PhysRevA.102.062408}, offering a complementary strategy for error mitigation. 
As detailed in Appendix~\!\ref{app:CWStark}, the numerical simulations reveal that such CW Stark drives can robustly suppress static ZZ couplings even in the large-detuning regime. Notably, the microwave phase of the CW drive enables dynamic control, allowing both suppression and activation of the ZZ interaction. This capability ensures compatibility with single-qubit operations while facilitating implementation of CPhase and CZ gates, presenting an effective approach for fixed-coupling architectures.
Separately, resonator-assisted protocols (e.g., Huang et al.~\cite{PhysRevApplied.22.034007}) offer another compatible approach for large-detuning systems. By leveraging resonator-induced phase interactions to fully neutralize static ZZ coupling, such schemes enable high-fidelity entangling gates including cross-resonance CNOT gates ($\sim$40 ns) and adiabatic CZ gates ($\sim$140 ns).



To probe the dynamics of the microwave-activated transition, we apply a flat-top Gaussian-edged pulse ($\sigma = 10$ ns, total edge duration = 4$\sigma$) to qubit $q_{\sss B}$. The pulse profile is illustrated in Fig.~\!\ref{fig:pulse}(a). When driving at the transition frequency ($\omega_d = \omega_{\sss S}$), the protocol achieves coherent states exchange between $|001\rangle$ and $|110\rangle$, as shown in Fig.~\!\ref{fig:pulse}(b). Meanwhile, all non-target computational states maintain negligible population, confirming the state selectivity of our protocol and validating the state-transfer due to the microwave-activated transition. To further characterize the microwave-activated protocol, we initialize the system in $|001\rangle$ and vary the drive frequency ($\omega_d$) and pulse duration. Figure ~\!\ref{fig:pulse}(c) illustrates the resulting the population of $|110\rangle$ near the predicted resonance $\omega_{\sss S} \sim \omega_{\sss B} + \Delta_{\sss AC}$, revealing a pronounced population transfer. The corresponding oscillation frequencies extracted from time-domain, as shown in Fig.~\!\ref{fig:pulse}(d), exhibit excellent agreement with the predicted value of approximately 1.25 MHz obatined from Eq.~\!\eqref{eq:eq004}. 
To further quantify the off-resonant transitions near $\omega_{\sss S}$, Figure~\!\ref{fig:pulse}(e) shows the oscillation frequencies of the $|001\rangle$ population across a wide drive-frequency range, initialized from $|001\rangle$. The spectrum exhibits a primary blue branch resonance (blue point), a red branch sideband (red point), sidebands near $\omega_{\sss B}$ from higher-order nonlinear transitions, with negligible distant features. Notably, the spectral sparsity around the target $\omega_{\sss S}$, inherent to the large-detuning regime ($\omega_{\sss B} \gg \omega_{\sss A}, \omega_{\sss C}$), suppresses parasitic couplings and ensures strong isolation from the nearest sidebands.

\begin{figure}[t]
	\includegraphics[width=1.0\linewidth]{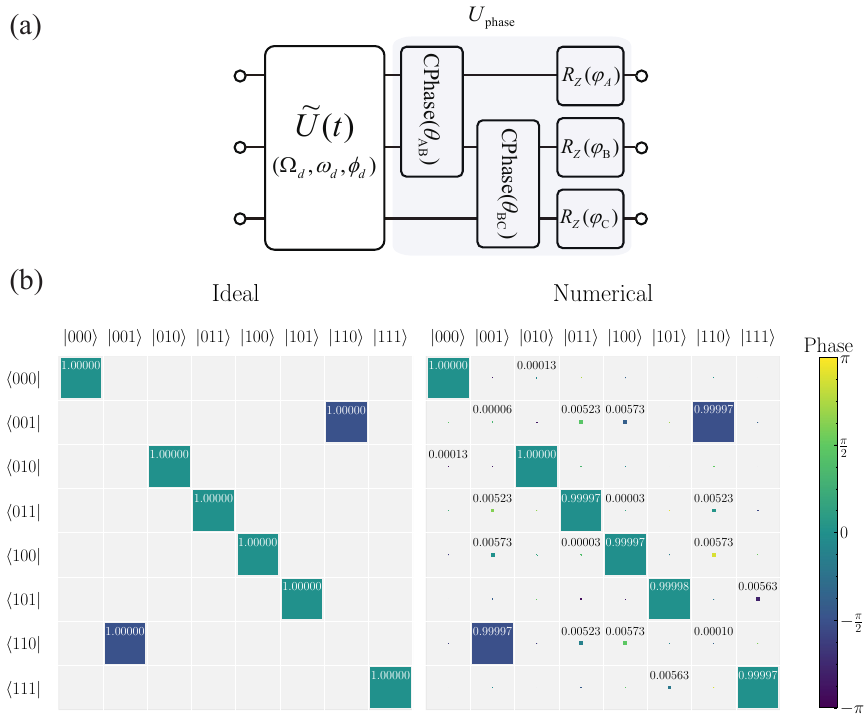}
	\caption{Phase correction protocol and truth table. (a) Schematic of the phase correction protocol, integrating the microwave-activated unitary $\widetilde{U}(\Omega_d, \omega_d, \phi_{d})$ and the correction unitary $U_{\text{phase}}$. In $\widetilde{U}(\Omega_d, \omega_d, \phi_{d})$,  $\phi_{d}$ represents the dynamically optimized pulse phase designed to suppress transient phase errors. The correction unitary $U_{\text{phase}}$, applied after the microwave-activated operation, consists of two CPhase terms ($\theta_{\sss AB}$, $\theta_{\sss BC}$) for qubit pairs $AB$ and $BC$, as well as three single-qubit virtual $Z$-rotations ($\varphi_{\sss A}, \varphi_{\sss B}, \varphi_{\sss C}$). By optimizing these six parameters ($\theta_{\sss AB}, \theta_{\sss BC}, \varphi_{\sss A}, \varphi_{\sss B}, \varphi_{\sss C}, \phi_{d}$), the protocol achieves high-fidelity three-qubit gates with minimized static and dynamic phase errors.
	(b) Truth table at $\Omega_d/2\pi \approx 90$ MHz, comparing the ideal gate unitary (left) with numerical simulation results (right). The protocol achieves an average fidelity of 99.9\% and demonstrates near-perfect population transfer.
	}
	\label{fig:Figs_unitary}
\end{figure}

\begin{figure*}[t]
	\centering
	{\includegraphics[width = 0.92\linewidth]{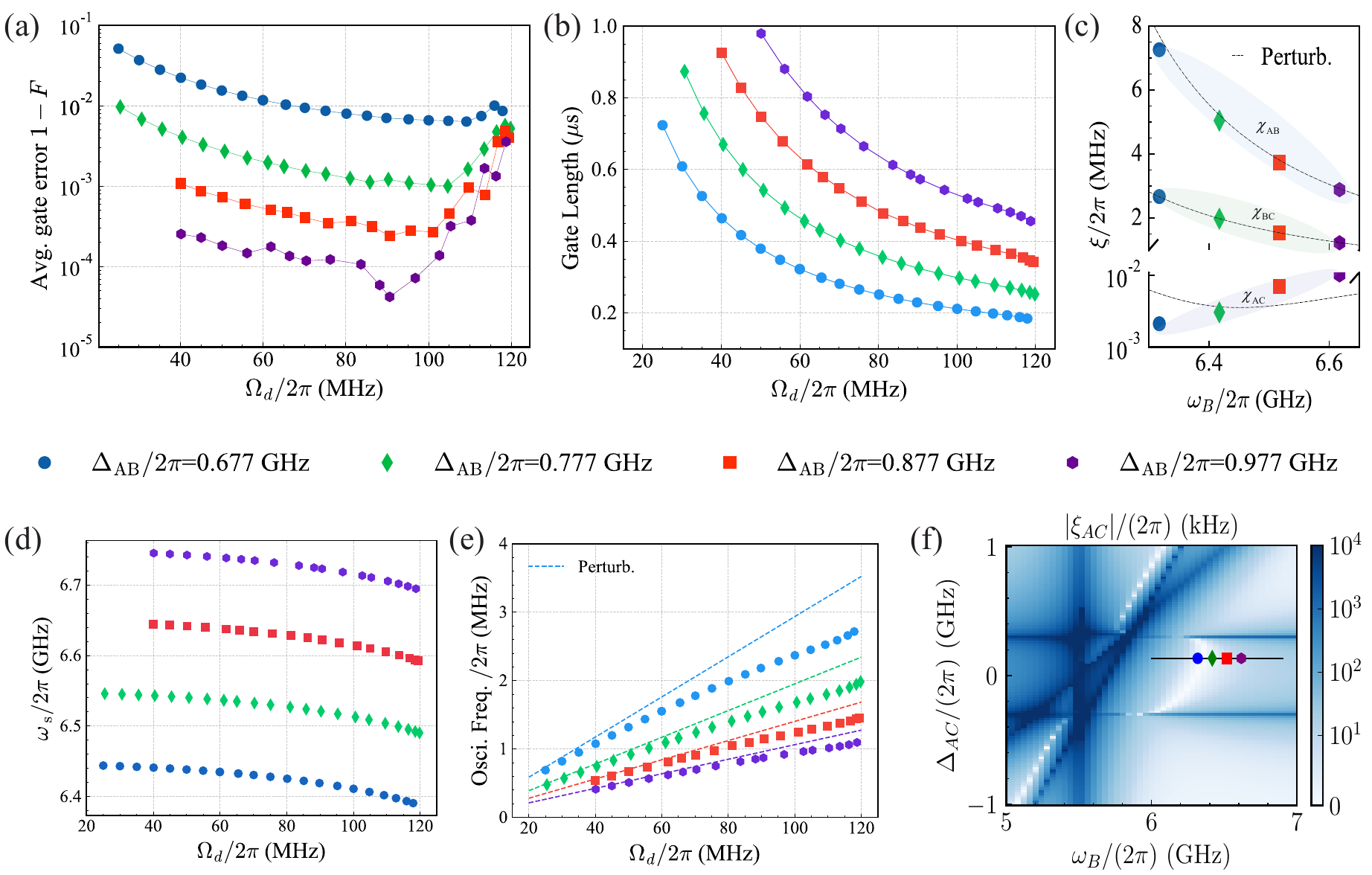}}
	\caption{Gate error versus drive amplitude and qubit detuning. (a) Average gate error $1 - F$ as a function of drive amplitude $\Omega_d$ for varying NN detunings $\Delta_{\sss AB}/2\pi = \{0.677, 0.777, 0.877, 0.977\}\,$GHz (blue, green, red, purple curves). Increasing  $\Delta_{\sss AB}$ reduces errors from $\sim 10^{-2}$ to $\sim 10^{-4}$, demonstrating the advantage of large detuning. (b) Gate duration decreases with increasing drive amplitude $\omega_d$ and decreasing $\Delta_{\sss AB}$, reflecting the inverse proportionality between oscillation frequency $\nu$ and $\Delta_{\sss AB}$ (Eq.~\!\eqref{eq:eq004}). (c) Static ZZ couplings $\xi_{\sss AB}$ and $\xi_{\sss BC}$ increase with reduced $\omega_{\sss B}$, while the NNN coupling $\xi_{\sss AC}$ remains suppressed below $10$ kHz due to the large detuning. Dashed lines correspond to fourth-order perturbative calculations of static ZZ couplings.
	(d) Transition frequency $\omega_{\mathrm{s}}$ as a function of $\Omega_d$, showing systematic shifts under varying drive amplitudes and detunings.  
	(e) Numerical results for $\nu$ (points) compared to perturbative predictions. Numerical results (points) deviate from perturbative predictions (solid line, Eq.~\!\eqref{eq:eq004}) at high $\Omega_d$, particularly for small detunings, highlighting the need for beyond-perturbative modeling in strongly driven regimes. (f) $|\xi_{\sss AC}|$ versus $\omega_{\sss B}$ and $\omega_{\sss A}$ (fixed $\omega_{\sss C}/2\pi = 5.507\,$GHz). Black solid-line traces highlight four selected NN detuning configurations, corresponding to $\Delta_{\sss AB}/2\pi = \{0.677, 0.777, 0.877, 0.977\}\,$GHz, concentrated within near-zero $\xi_{\scriptscriptstyle AC}$ regions, revealing strong long-range ZZ suppression in the large-detuning regime.} 
	\label{fig:Figs_Fidelity_wB}
\end{figure*}

\section{microwave-activated three-qubit Gate performance\label{sec:Gateperformance}}

The microwave-activated state-transfer mechanism enables coherent population exchange between states $|001\rangle$ and $|110\rangle$ in a three-qubit system, providing the foundation for an three-qubit gate. During this process, the target states acquire a characteristic $-i$ phase factor while all other computational states remain unaffected. However, the drive pulse inevitably induces off-resonant transitions that dynamically perturb neighboring states, leading to unintended phase shifts. Additionally, static ZZ couplings inherent to fixed-frequency architectures cause conditional phase accumulations, further distorting the ideal unitary operation. 
To mitigate these undesired effects, we implement a targeted phase correction protocol using the unitary $U_{\mathrm{phase}}$, combining CPhase terms and single-qubit $Z$-rotations, as illustrated in Fig.~\!\ref{fig:Figs_unitary}(a). The CPhase operation is defined by the diagonal matrix, $U_{\mathrm{CPhase}} = \mathrm{diag}\left[1, 1, 1, e^{-i\theta_{\sss BC}}, 1, 1, e^{-i\theta_{\sss AB}}, e^{-i(\theta_{\sss AB}+\theta_{\sss BC})}\right]$, where $\theta_{jk} = \xi_{jk} t$ encodes the ZZ-mediated phase accumulation between NN qubits $q_j$ and $q_k$, while single-qubit phases ($\varphi_{\sss A},~\varphi_{\sss B},~\varphi_{\sss C}$) can be corrected using virtual Z gates~\!\cite{McKay2017}. 
Additionally, the microwave pulse phase $\phi_{d}$ is also considered to ensure full cancellation of both dynamic and static phase errors. This hierarchical compensation scheme addresses dominant CPhase errors between NN qubits through tailored corrections while adhering to hardware constraints that restrict high-fidelity CPhase operations to directly coupled pairs. Residual errors from NNN ZZ coupling (e.g., $\xi_{\sss AC} \approx 10$ kHz) are passively suppressed by the intrinsic dispersive regime, eliminating the need for additional error suppression. By canceling both static and dynamically induced phase errors without compromising experimental simplicity, the protocol can achieve robust gate performance in scalable fixed-frequency qubit systems.

By applying $U_{\mathrm{phase}}$ to the microwave-activated unitary $\widetilde{U}$, the desired three-qubit gate unitary $U^{\prime}$ can be expressed as
\begin{equation}\label{eq:eq005}
U^{\prime} = U_{\mathrm{phase}} \widetilde{U},
\end{equation}
where the corresponding ideal matrix in the computational basis is
\begin{equation}\label{eq:eq006}
U_{i}=\left(\begin{array}{cccccccc}
1 & 0 & 0 & 0 & 0 & 0 & 0 & 0 \\
0 & 0 & 0 & 0 & 0 & 0 & -i & 0 \\
0 & 0 & 1 & 0 & 0 & 0 & 0 & 0 \\
0 & 0 & 0 & 1 & 0 & 0 & 0 & 0 \\
0 & 0 & 0 & 0 & 1 & 0 & 0 & 0 \\
0 & 0 & 0 & 0 & 0 & 1 & 0 & 0 \\
0 & -i & 0 & 0 & 0 & 0 & 0 & 0 \\
0 & 0 & 0 & 0 & 0 & 0 & 0 & 1
\end{array}\right),
\end{equation}
which induces the $|001\rangle \leftrightarrow |110\rangle$ exchange with $-i$ phase, analogous to an iSWAP operation scaled to three qubits. This operation ensures that only the targeted states undergo the desired phase transformation, while all other computational states remain unaffected.

\begin{figure}[t]
	\includegraphics[width=0.47\textwidth]{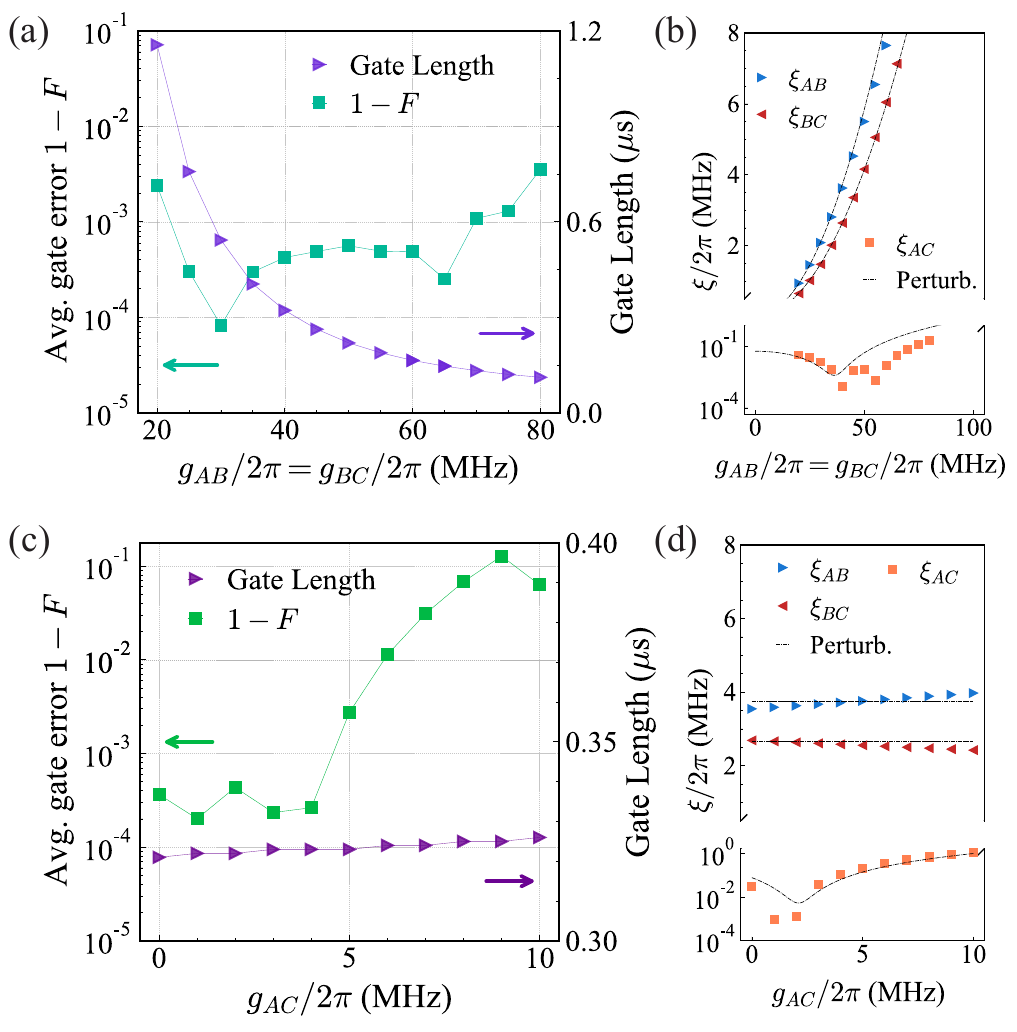}
	\caption{Gate error dependence on qubit coupling parameters. (a) Gate error (cyan curve, left axis) and gate duration (purple curve, right axis) as functions of the NN coupling strengths $g_{\sss AB}/2\pi = g_{\sss BC}/2\pi$, under a fixed microwave drive amplitude $\Omega_d = 100$ MHz. The gate error remains below $10^{-3}$ across a broad NN coupling range of 40–60 MHz, demonstrating robust performance. Gate durations exhibit an inverse proportionality to coupling strength, while gate durations inversely scale with coupling strength, requiring longer operation times for weaker couplings. (b) Static ZZ couplings ($\xi_{\sss AB}, \xi_{\sss BC}, \xi_{\sss AC}$) as functions of $g_{\sss AB}/2\pi = g_{\sss BC}/2\pi$, with dashed curves showing perturbative predictions. (c) Gate error (green curve) and duration (purple curve) versus NNN coupling $g_{\sss AC}/2\pi$ under fixed $g_{\sss AB}/2\pi = g_{\sss BC}/2\pi = 40$ MHz and $\Omega_d = 100$ MHz. Error suppression below $10^{-3}$ persists for $g_{\sss AC}/2\pi \leq 4$ MHz but degrades with stronger couplings. (d) Static ZZ couplings ($\xi_{\sss AB}, \xi_{\sss BC}, \xi_{\sss AC}$) as functions of $g_{\sss AC}/2\pi = g_{\sss BC}/2\pi$, with dashed curves showing perturbative predictions. Growth of NNN static ZZ coupling $\xi_{\sss AC}$ as a function of $g_{\sss AC}/2\pi$, correlating with the fidelity degradation observed in (c). }
	\label{fig:Figs_Fidelity_gNN}
\end{figure}

To determine the optimal operating parameters for a three-qubit gate, the numerical optimization proceeds through four key stages. The process begins by conducting 2D Rabi spectroscopy at a fixed driving strength $\Omega_d$ to accurately determine the dressed transition frequency $\omega_{\sss S}$. Based on this, we perform numerical simulations of the system dynamics to identify the minimal gate duration required to enable a complete $\ket{001}\leftrightarrow\ket{110}$ state exchange.
We then optimize the gate parameters by performing a local search in the neighborhood of the initial parameter space to maximize state-transfer population. This optimization incorporates the quadrature term (derivative removal via adiabatic gate, DRAG) for pulse edges to suppress leakage errors~\!\cite{PhysRevLett.103.110501,PhysRevA.83.012308,PhysRevA.88.052330,PRXQuantum.1.020318}. Based on this, the microwave-activated unitary operation $\widetilde{U}$ is subsequently obtained from numerical simulations of the dynamical evolution  governed by Eq.~\!\eqref{eq:eq003}. After applying phase compensation to eliminate phase distortions, we construct the phase corrected unitary $ U^{\prime}$ using Eq.~\!\eqref{eq:eq005}. The average gate fidelity is evaluated using the standard metric~\!\cite{PhysRevB.85.241401,Phys.Lett.A367.47,PhysRevB.96.024504},
\begin{equation}
F = \frac{\left|\mathrm{Tr}(U_{\mathrm{ideal}}^{\dagger} U^{\prime})\right|^{2} + \mathrm{Tr}\left(U^{\prime \dagger} U^{\prime} \right)}{d(d + 1)},
\end{equation}
where $d = 8$ represents the total dimension for the three-qubit system and $U_{\mathrm{ideal}}$ denotes the ideal gate defined in Eq.~\!\eqref{eq:eq006}.

Following the targeted phase correction protocol applied to the derived evolution operator, we achieve an average gate fidelity exceeding $\geq 99.9\%$ under parameters ($\Omega_d/2\pi \approx 90$~MHz, gate duration $\approx 400$~ns). Figure~\!\ref{fig:Figs_unitary}(b) compares the ideal truth table of the target operation $ U_{\mathrm{ideal}}$ (left panel) with the numerically corrected unitary $U^\prime$ (right panel). The comparison demonstrates consistent population transfer across all computational states and strong agreement between the theoretical gate operation and simulated outcomes.

Next, we investigate the gate performance under varying qubit parameters by exploring its dependence on NN detunings. We fix the frequencies of qubits $q_{\sss A}$ and $q_{\sss C}$ at 5.641 GHz and 5.507 GHz, while varying the frequency of $q_{\sss B}$ across $\omega_{\sss B}/2\pi = \{6.317, 6.417, 6.517, 6.617\}\,$GHz. This corresponds to NN detunings $\Delta_{\sss AB}/2\pi = \{0.677, 0.777, 0.877, 0.977\}$ GHz. Numerical simulations under these conditions are summarized in Fig.~\!\ref{fig:Figs_Fidelity_wB}. Figure~\!\ref{fig:Figs_Fidelity_wB}(a) shows the average gate error $1 - F$ as a function of drive amplitude $\Omega_d$ for different $\Delta_{\sss AB}$. The blue, green, red, and purple curves correspond to $\Delta_{\sss AB}/2\pi = \{0.677, 0.777, 0.877, 0.977\}$~GHz, respectively. Notably, increasing $\Delta_{\sss AB}$ enhances gate fidelity, with gate errors decreasing from around $10^{-2}$ at $\Delta_{\sss AB} = 0.677 $ GHz to around $10^{-4}$ at $\Delta_{\sss AB} = 0.977 $ GHz. This improvement highlights the advantage of operating in the large-detuning regime.  Figures~\!\ref{fig:Figs_Fidelity_wB}(b)-(e) present the numerical results for gate duration, transition frequency $\omega_{\mathrm{s}}$, and oscillation frequency $\nu$ as functions of $\Omega_d$. As shown in Fig.~\!\ref{fig:Figs_Fidelity_wB}(b), the gate duration decreases with higher drive amplitudes and smaller NN detunings, reflecting the inverse proportionality between $\nu$ and $\Delta_{\sss AB}$, predicted by Eq.~\!\eqref{eq:eq004}. This relationship is further corroborated in Fig.~\!\ref{fig:Figs_Fidelity_wB}(e), where the solid line represents the perturbative results from Eq.~\!\eqref{eq:eq004}. Deviations from perturbation theory emerge at larger drive amplitudes, particularly for smaller $\Delta_{\sss AB}$, indicating the restricted validity of the perturbative approximation under strong driving. The static ZZ coupling characteristics are systematically investigated in Fig.~\!\ref{fig:Figs_Fidelity_wB}(c). While NN static ZZ couplings ($\xi_{\sss AB}$, $\xi_{\sss BC}$) grow with decreasing detuning $\Delta_{\sss AB}$, the NNN coupling $\xi_{\sss AC}$ remains suppressed below $10$ kHz, demonstrating the protective effect of large NN detunings. The dashed curves confirm agreement with perturbative calculations, where NN ZZ couplings follow the analytical expression in Eq.~\eqref{eq:eqZZNNij}, while the NNN coupling is derived from fourth-order perturbation theory, as detailed in Refs.~\!\cite{PhysRevX.11.021058,PhysRevApplied.16.054020}.
Figure~\ref{fig:Figs_Fidelity_wB}(f) shows the NNN ZZ coupling $\xi_{\sss {AC}}$ as a function of $\omega_{\sss B}$ and $\Delta_{\sss AC}$, with the colored points along black solid line marks four configurations of the selected NN detunings $\Delta_{\sss AB}/2\pi = \{0.677, 0.777, 0.877, 0.977\}$~GHz. These configurations consistently reside in parameter regimes where $\xi_{\scriptscriptstyle AC}$ approaches zero, reflecting the inherent long-range ZZ suppression enabled by the large-detuning regime.

We further investigate the gate performance under varying qubit parameters, including NN coupling strengths $g_{\sss AB}, g_{\sss BC}$ and NNN coupling $g_{\sss AC}$, with $\omega_{\sss B}$ fixed at $6.517\,$GHz. Figure~\!\ref{fig:Figs_Fidelity_gNN}(a) shows the gate error (cyan curve) as a function of $g_{\sss AB} = g_{\sss BC}$ under a fixed drive amplitude $\Omega_d = 100$~MHz. Notably, the gate achieves sub-$10^{-3}$ errors across a broad NN coupling range of $[40, 60]\,$MHz. The corresponding gate durations (purple curve) exhibit an inverse relationship with coupling strength, where weaker NN couplings necessitate longer gate times. The static ZZ coupling are analyzed in Fig.~\!\ref{fig:Figs_Fidelity_gNN}(b). Reducing NN coupling strengths suppresses the NN static ZZ couplings ($\xi_{\sss AB}, \xi_{\sss BC}$), while stronger NN couplings amplify the NNN coupling $\xi_{\sss AC}$, leading to long-range crosstalks and phase errors. We further explore the gate performance under varying NNN coupling $g_{\sss AC}$ with $g_{\sss AB}/2\pi = g_{\sss BC}/2\pi = 40$ MHz and $\Omega_d = 100$ MHz, as shown in Fig.~\!\ref{fig:Figs_Fidelity_gNN}(c). We further investigate the influence of NNN coupling $g_{\sss AC}$ on gate performance under fixed parameters $g_{\sss AB}/2\pi = g_{\sss BC}/2\pi = 40$ MHz and drive amplitude $\Omega_d = 100$ MHz. As shown in Fig.~\!\ref{fig:Figs_Fidelity_gNN}(c), the gate error remains below $10^{-3}$ for $g_{\sss AC}/2\pi \in [0, 4]$ MHz, but increases as $g_{\sss AC}$ exceeds this range. This degradation correlates with the rise of static ZZ coupling $\xi_{\sss AC}$, as demonstrated by its growth with $g_{\sss AC}$ in Fig.~\!\ref{fig:Figs_Fidelity_gNN}(d). Notably, we observe a pronounced correlation between gate error and the NNN static ZZ coupling $\xi_{\sss AC}$. The underlying mechanism stems from the persistent static ZZ coupling in fixed-coupling architectures, which introduces undesired CPhase errors. While NN ZZ couplings ($\xi_{\sss AB}$ and $\xi_{\sss BC}$) have been actively compensated through phase compensation protocols, the NNN static ZZ coupling $\xi_{\sss AC}$ induces uncompensated phase errors. These residual errors accumulate during gate operations, ultimately limiting the gate fidelity as $g_{\sss AC}$ increases.

We further evaluate gate performance through process fidelity simulations (Appendix~\!\ref{app:ProcessFidelity}), which demonstrate high-fidelity performance across various operating conditions. In the absence of decoherence, process fidelity exceeds 98.5\% at $\Delta_{\mathrm{AB}} = 0.877$ GHz and reaches above 99.0\% at $\Delta_{\mathrm{AB}} = 0.977$ GHz. When incorporating realistic decoherence with $T_1 = T_2^* = 200\ \mu\text{s}$, $100\ \mu\text{s}$, and $50\ \mu\text{s}$, the fidelity remains near 98\%,  97\%, and  96\%, respectively. This demonstrates compatible compatibility with state-of-the-art coherence times ($>200$ $\mu$s in fixed-frequency transmons~\cite{Place2021.0.3ms,Wang2022.0.5ms,Bal2024,interfacialdielectricloss2024,Bu2025,Tuokkola2025T11ms}).
Notably, all-microwave CW Stark drives can be used to effectively suppress static ZZ couplings in the large-detuning regime essential to our system (Appendix~\ref{app:CWStark}).  The microwave phase of the CW drive provides dynamic control over the ZZ interaction, permitting both suppression for high-fidelity single-qubit operations and activation for controlled-phase implementations (high-fidelity CPhase and CZ gates). This approach achieves CZ gate fidelities exceeding 99.43\% (gate length ~200 ns) as reported in literature~\cite{PhysRevLett.127.200502,PhysRevLett.129.060501}.
In parallel, Huang et al.~\cite{PhysRevApplied.22.034007} recently introduced  an effective, high-performance solution for suppressing static ZZ coupling in large-detuning fixed-frequency architectures, achieving simulated gate fidelities (40-ns CNOT and 140-ns adiabatic CZ gate) exceeding 99.99\%.
These adaptability positions the three-qubit gate framework as a practical solution for high-fidelity multi-qubit operations in real-world quantum processors, combining high fidelity, decoherence resilience, dynamic control, and scalability.

 \begin{figure}[t]
   \includegraphics[width=0.45\textwidth]{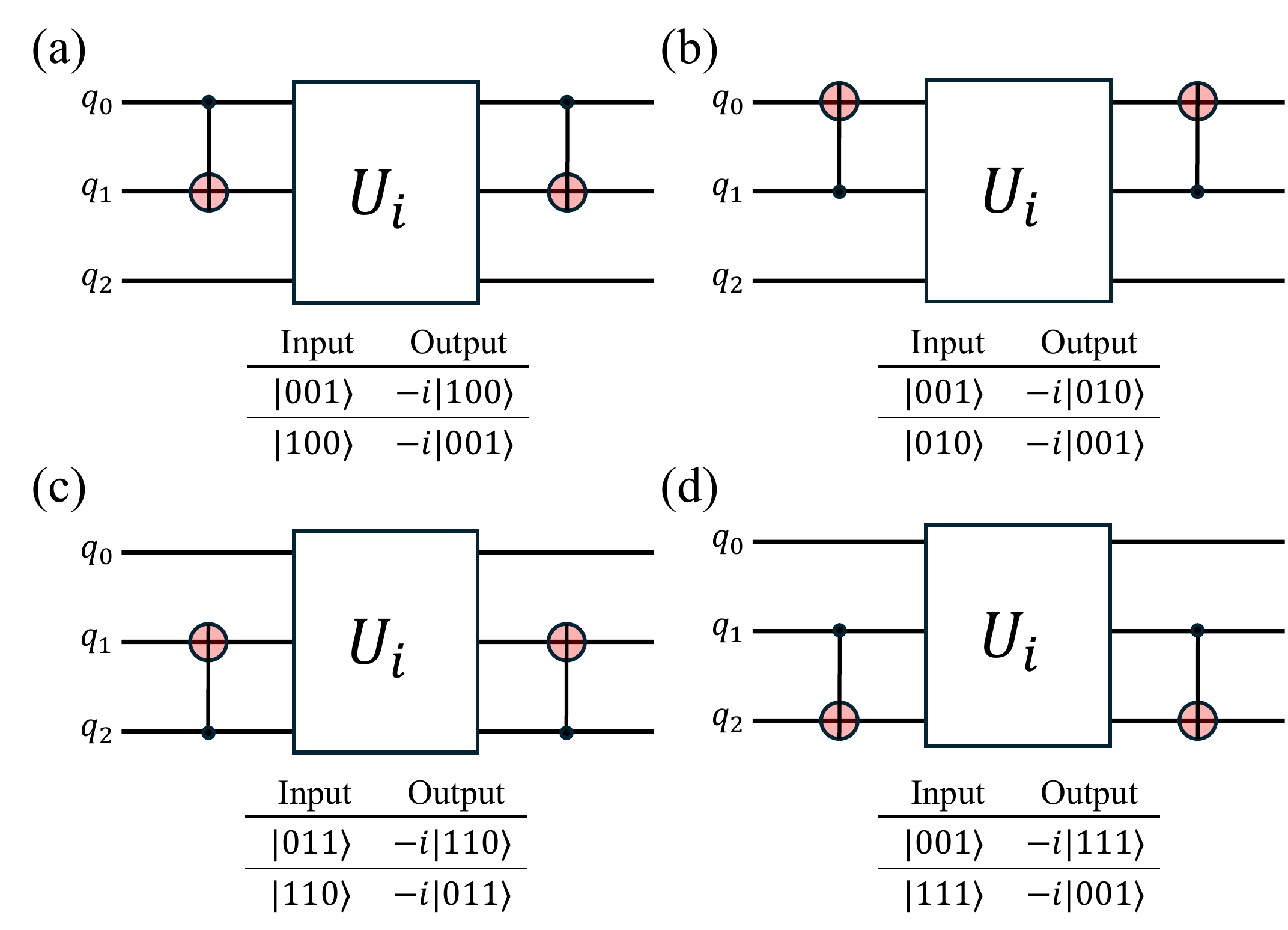}
    \caption{
    Implementation of iFredkin (controlled-SWAP) gate through combing $U_i$ and CNOT operations.
    (a) $\ket{0}$-control on $q_2$: swaps $q_1$ and $q_3$ ($\ket{001}\leftrightarrow\ket{110}$) with $-i$ phase.  
    (b) $\ket{0}$-control on $q_1$: swaps $q_2$ and $q_3$ ($\ket{001}\leftrightarrow\ket{010}$).  
    (c) $\ket{1}$-control on $q_2$: swaps $q_2$ and $q_3$ ($\ket{011}\leftrightarrow\ket{110}$).  
    (d) $\ket{1}$-control on $q_3$: global flip when $q_1=q_2$ ($\ket{001}\leftrightarrow\ket{111}$). Each configuration selectively swaps qubit pairs under distinct control conditions while preserving non-target states.}
      \label{fig:MASCNOT}
\end{figure}

The microwave-activated three-qubit gate proposed here can be used to directly constructing universal iFredkin (controlled-SWAP) gates. These gates share a common operational framework as members of the controlled-SWAP family, where the iFredkin gate requires only two additional controlled-NOT (CNOT) gate operations to transform the base unitary. Figure~\!\ref{fig:MASCNOT} demonstrates the direct construction of an iFredkin gate using the microwave-activated three-qubit gate. Specifically, applying CNOT gates on NN qubits enables different iFredkin gate configurations, as shown in Fig.~\!\ref{fig:MASCNOT}. 
By leveraging high-fidelity all-microwave two-qubit gates in existing large-detuning fixed-frequency architectures~\cite{PhysRevApplied.14.044039,PhysRevApplied.22.034007}, such as CNOT gates and CZ gates, our gate scheme enables the implementation of the iFredkin gate without requiring any hardware modifications.

\section{Conclusion\label{sec:Conclusion}}

In summary, we demonstrate a microwave-activated three-qubit gate scheme designed for large-detuning fixed-frequency transmon architectures, offering a practical approach to extend the operational capabilities of superconducting quantum processors. By virtue of the intrinsic third-order nonlinearity of transmon qubits under large-detuning conditions ($|\Delta| \gg g$), we implement a three-qubit operation that coherently exchanges $\ket{001} \leftrightarrow \ket{110}$ states via a single microwave drive. Numerical simulations confirm $99.9\%$ average gate fidelity under large-detuning conditions.
The protocol maintains high performance under decoherence while exhibiting intrinsic resilience to fabrication-induced parameter variations. Additionally, it demonstrates compatibility with high-fidelity two-qubit gate frameworks in fixed-frequency architectures, simultaneously preserving spectral isolation to mitigate frequency crowding. 

The proposed three-qubit gate scheme eliminates tunable elements to avoid decoherence pathways associated with flux control circuitry, while utilizing large-detuning operation to simultaneously achieve natural scalability and improved robustness against parameter variations.
By operating in the dispersive coupling regime, the protocol not only suppresses unwanted interactions but also simplifies control requirements through all-microwave driving. This enables direct implement of three-qubit gates without decomposing them into elementary single- and two-qubit operations, which reduces circuit depth and enhances algorithmic flexibility for near-term quantum applications. Our results reinforce the potential of fixed-frequency superconducting platforms as a viable substrate for advancing scalable quantum computation, bridging the gap between robust two-qubit operations and the demands of multi-qubit control.

\begin{acknowledgments}
This work was supported by National Natural Science Foundation of China (Grants Nos.~\!92265207, T2121001, 12504593, 12122504, T2322030, 92365301, and 12404578), the Innovation Program for Quantum Science and Technology (Grant No.~\!2021ZD0301800), the China Postdoctoral Science Foundation (Grant~No.~\!GZB20240815), Beijing Nova Program (Nos.~\!20220484121, 2022000216), and Beijing National Laboratory for Condensed Matter Physics (2024BNLCMPKF022).
\end{acknowledgments}

\appendix

\section{The Effective Hamiltonian \label{app:4WM}}

\begin{figure}[!htbp]
	\centering 
	\includegraphics[width=0.36\textwidth]{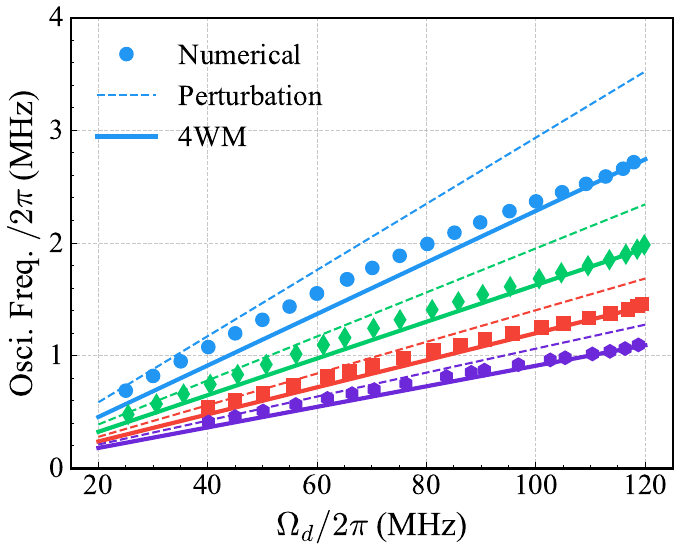}
	\caption{Comparison of oscillation frequencies between perturbative predictions (dashed lines, Eq.~\!\eqref{eq:eq004}) and results derived from the four-wave mixing process (solid lines, Eq.~\!\eqref{eq:eq004-4WM}).}
	\label{fig:Figs_4MW}
\end{figure}

In this section, we briefly discuss the physical origin of four-wave mixing (4WM) in our system. This phenomenon arises primarily from the inherent nonlinearity of the central qubit $ q_{\sss B} $, consistent with mechanisms reported in Refs.~\!\cite{science.aaa2085two-photon,PhysRevX.10.021038,PhysRevA.95.063848}. To clarify, we approximate the system as three linear photonic modes coupled via the Josephson junction of $ q_{\sss B} $, governed by the Hamiltonian
\begin{equation}
    \hat{H} = \sum_{j} \omega_j \hat{a}_j^\dagger \hat{a}_j - E_J  \cos\hat{\varphi} +  \Omega_d \cos{(\omega_d{t})}(\hat{a}^\dagger_{\sss B} + \hat{a}_{\sss B}),
\end{equation}
where $j \in {A, B, C}$, $E_J$ denotes the Josephson energy, and the phase operator $\hat{\varphi} = \sum_{\sss j} \varphi_j (\hat{a}_j + \hat{a}_j^\dagger)$ represents the total phase across the junction. This decomposes into a linear combination of phase contributions from individual modes, with $\varphi_j$ quantifying the zero-point phase fluctuation for mode $j$.
 The quadratic order contribution to the cosine potential, which encodes the linear mode hybridization, can be fully incorporated within the dressed mode. Consequently, the dominant nonlinear coupling arises from quartic interactions. By systematically neglecting sixth- and higher-order terms, we obtain the approximate Hamiltonian,
\begin{equation}
\hat{H}^{(0)} \approx \sum_{j} \tilde{\omega}_j \hat{a}_j^\dagger \hat{a}_j - \frac{E_J}{24} \hat{\varphi}^4 + \Omega_d \cos{(\omega_d{t})}(\hat{a}^\dagger_{\sss B} + \hat{a}{\sss B}).
\end{equation}
To eliminate the explicit time dependence, we transition to a frame rotating at the drive frequency $\omega_d$ while systematically discarding counter-rotating terms,
\begin{equation}
    \hat{H}^{(1)} \approx \sum_{j} \delta_j \hat{a}_j^\dagger \hat{a}_j - \frac{E_J}{24} \hat{\varphi}^4 +  \Omega_d(\hat{a}^\dagger_{\sss B} + \hat{a}_{\sss B}),
\end{equation}
where  $\delta_j = \tilde{\omega}_j - \omega_d$.
The linear drive term can be subsequently eliminated by moving to a displaced frame, approximated by its mean-field amplitude $\tilde{\xi}_d = \xi_d e^{-i\omega_d t}$, with $\xi_d = \Omega_d / (\omega_d - \omega_{\sss B})$ ~\!\cite{science.aaa2085two-photon,science.aat3996,Leghtas2015,ZhaoPengPhysRevApplied.10.024019}
, capturing drive-induced displacement. 
In the displaced frame with respect to the unitary transformation,
\begin{equation}
U(t) = e^{-\tilde{\xi}_d \hat{a}{\sss B}^\dagger + \tilde{\xi}_d^* \hat{a}{\sss B}},
\end{equation}
which explicitly shifts the annihilation operator as $U(t)\hat{a}{\sss B} U(t)^{-1} = \hat{a}_{\sss B} + \tilde{\xi}_d$. Applying this displacement transformation to $\hat{H}^{(1)}$ then generates the displaced Hamiltonian, 
\begin{equation}
    \hat{H}^{(2)} \approx \sum_{j} \delta_j \hat{a}_j^\dagger \hat{a}_j - \frac{E_J}{24} \hat{\varphi^{\prime}}^4,
\end{equation}
where the displaced phase variable is now given by,
\begin{equation}
\begin{aligned}
\hat{\varphi}^{\prime} =\ \varphi_{\sss B} \left( \tilde{\xi}_d  + \tilde{\xi}_d^* \right)
+ \sum_{ \sss j \in \{A,B,C\}} \varphi_j \left( \hat{a}_{\sss j}  + \hat{a}_j^\dagger \right),
\end{aligned}
\end{equation} 
Finally, expanding the quartic nonlinearity $\hat{\varphi}^{\prime 4}$
and retaining only the dominant contributions leaves us with an effective nonlinear Hamiltonian characterized by three key processes,
\color{black}
\begin{equation}
\hat{H}^\prime \approx \hat{H}_{\text{Stark}} + \hat{H}_{\text{Kerr}} + \hat{H}_{\text{4WM}},
\end{equation} 
which includes three dominant contributions: Stark shifts, Kerr nonlinearities, and four-wave mixing. The Stark shift term modifies mode frequencies, i.e.,
\begin{equation}
    \hat{H}_{\text{Stark}}\!=\!\left(\delta_{\sss B} \!-\! 2 \chi_{\sss BB} \left|\xi_d\right|^2\right) \hat{a}_{\sss B}^\dagger \hat{a}_{\sss B} +\!\!\sum_{\sss j \in \{A,C\}} \!\!(\delta_j - \chi_{Bj} |\xi_d|^2) \hat{a}_j^\dagger \hat{a}_j,
\end{equation} 
while Kerr nonlinearities introduce anharmonicity via term
\begin{equation}
\begin{aligned}
    \hat{H}_{\text{Kerr}} =&  -\sum_{\sss j = A,B,C} {\chi_{jj}} \hat{a}_{j}^{\dagger 2} \hat{a}_{j}^2 - \sum_{\sss  k,l} 
\chi_{kl} \hat{a}_k^\dagger \hat{a}_k \hat{a}_l^\dagger \hat{a}_l,
\end{aligned}
\end{equation} 
with coefficient $\chi_{kl} = E_J \phi_k^2 \phi_l^2 $ and $kl\in \{AB, BC, AC\}$. The critical 4WM term emerges as  
\begin{equation}
\hat{H}_{\text{4WM}} = g_{3} \hat{a}_{\sss A}^\dagger \hat{a}_{\sss B}^\dagger \hat{a}_{\sss C} + \text{H.c.},
\end{equation}  
and the third-order nonlinear coupling $g_3$ is given by
\begin{equation}\label{eq:eq004-4WM}
    g_{3} = -\xi_d \sqrt{\chi_{\sss AB} \chi_{ \sss BC}},
\end{equation}
where $ \chi_{\sss AB} $ corresponds to the dispersive shift of  qubit $ B $, and it can be approximated as  
\begin{equation}
\chi_{\sss AB} \approx \frac{1}{1 + \Delta_{\sss AB}/\alpha_{\sss B}}\left(\frac{g_{\sss AB}}{\Delta_{\sss AB}}\right)^2.
\end{equation}  
The analytical expression derived here exhibits significantly improved agreement with experimental data compared to the perturbative approach outlined in Eq.~\!\eqref{eq:eq004}, as quantitatively demonstrated in Fig.~\!\ref{fig:Figs_4MW}.  The 4WM process originates from the qubit-mediated nonlinear interaction among qubit modes where the drive photon parametrically activates the resonant coupling condition  
\begin{equation}
    \omega_d = \omega_{\sss B} \!+\! \omega_{\sss C} \!-\! \omega_{\sss A} \!-\! \chi_{\sss BC} \!-\! |\xi_d|^2 (2\chi_{\sss BB} \!+\! \chi_{\sss BC} \!-\! \chi_{\sss AB}),
\end{equation}
where $\chi_{\sss BB} = \alpha_{\sss B} $. The above parametric resonance drives coherent population transfer between $\ket{001}$ and $\ket{110}$. A comprehensive perturbative analysis of this process  can found in Ref.~\!\cite{PhysRevLett.130.260601}. Such controlled nonlinear coupling forms the foundation of the microwave-activated three-qubit gate, directly linking the drive amplitude to the tunable 4WM strength, thereby enabling programmable entanglement operations.  

\section{Validity of the Rotating Wave Approximation}
\label{sec:RWA_validity}

This section numerically examines the validity of the rotating wave approximation (RWA) in the simulations. The total Hamiltonian in the lab frame is given by
\begin{equation}\label{eq:Hamiltonian_Full}
	\hat{H}_{\text{full}} = \hat{H}_0 + \hat{H}_d,
\end{equation}
with
\begin{equation}
\begin{aligned}
    \hat{H}_0 =& \sum_i \left(\omega_i \hat{a}^\dagger_i \hat{a}_i + \frac{\alpha_i}{2}\hat{a}^\dagger_i\hat{a}^\dagger_i\hat{a}_i\hat{a}_i\right) + \\& \sum_{i,j} g_{ij}(\hat{a}_i+\hat{a}^\dagger_i)(\hat{a}_j+\hat{a}^\dagger_j),   
\end{aligned}
\end{equation}
\begin{equation}
    \hat{H}_d = \Omega_d(t) \cos(\omega_d t + \phi_{d}) (\hat{a}_B^\dagger + \hat{a}_B).
\end{equation}
When transforming to the rotating frame with drive frequency $\omega_d$ via the unitary operator $\hat{U} = \exp\left(-i \omega_d t \sum_i \hat{a}_i^\dagger \hat{a}_i \right)$, the rotating-frame Hamiltonian becomes,
\begin{equation}\label{eq:Hamiltonian_FullRWA}
    \begin{aligned}
\hat{H}_{\mathrm{RF}} = & {\sum_i \left( (\omega_i - \omega_d) \hat{a}_i^\dagger \hat{a}_i + \frac{\alpha_i}{2} \hat{a}_i^\dagger \hat{a}_i^\dagger \hat{a}_i \hat{a}_i \right)} \\
& + \sum_{i,j} g_{ij} \left(\hat{a}_i \hat{a}_j^\dagger + e^{-i2\omega_d t} \hat{a}_i \hat{a}_j + h.c \right) \\
& + \frac{\Omega_d(t)}{2} \left(e^{i\phi_{d}} \hat{a}_{\sss _{\mathrm{\sss B}}} + e^{-i(2\omega_d t + \phi_{d})} \hat{a}_{\sss _{\mathrm{\sss B}}} + h.c \right).
\end{aligned}
\end{equation}
Note that the rotating wave approximation requires $g_{ij} \ll 2\omega_d$ and $\Omega_d \ll 4\omega_d$, thus the counter-rotating interaction terms ($e^{\pm i2\omega_d t} \hat{a}_i \hat{a}_j$) and driving terms ($e^{\pm i(2\omega_d t + \phi_{d})} \hat{a}_B^{(\dagger)}$) can be eliminated~\!\cite{PhysRevA.109.023703}.  Our parameters satisfy these conditions with substantial margin, qubit couplings ($g_{ij} = 20\sim{40}$ MHz), drive amplitudes ($\Omega_d = 40\sim{100}$ MHz), and drive frequencies ($\omega_d = 4\sim{7}$ GHz) yield maximum ratios $\max(g_{ij})/(2\omega_d) = 0.005$ and $\max(\Omega_d)/(4\omega_d) = 0.00625$. Counter-rotating terms oscillate at $2\omega_d \approx 8\sim{14}$ GHz (periods 71$\sim$125 ps), averaging to $<$0.63\% residuals during 400 ns gates ($>$ 3200 oscillation cycles).

\begin{figure}[!htbp]
    \centering
	\includegraphics[width=0.75\linewidth]{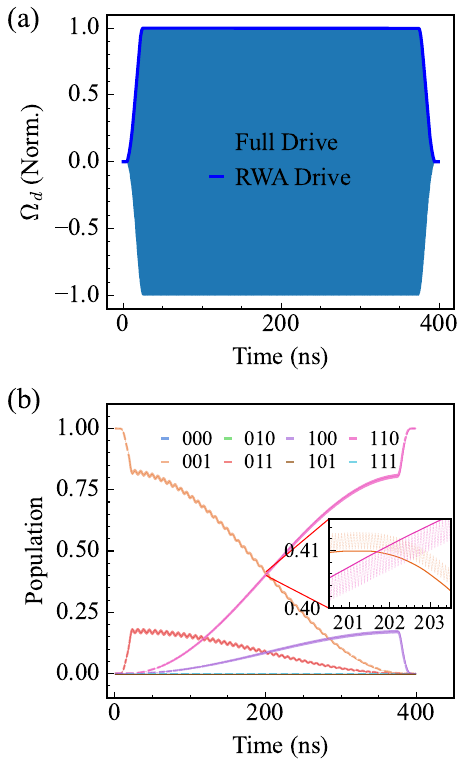}
	\caption{Comparison of full lab-frame and RWA Hamiltonian simulations for three-qubit system. (a) Microwave drive pulse: the full waveform (blue) explicitly includes the carrier frequency $\omega_d = 6.61347 $ GHz, while the RWA approximation (deep blue) retains only the pulse envelope, with the amplitude $\sim$ 100 MHz. (b) State population dynamics. Simulations using the full Hamiltonian (Eq.~\!\eqref{eq:Hamiltonian_Full}, dashed lines) and RWA Hamiltonian (Eq.~\!\eqref{eq:eq003}, solid lines). The full simulation exhibits minor high-frequency oscillations ($\sim$ 75.4 ps) from counter-rotating terms, causing only negligible transient deviations below 0.5\% in populations. Despite these oscillations, gate fidelity differences remain below $10^{-4}$, confirming the validity of the RWA approximation. Parameters identical to those in Fig.~\!\ref{fig:pulse}(b) were employed.}
    \label{fig:Figs_Pulse_RWA}
\end{figure}

Fig.~\ref{fig:Figs_Pulse_RWA} compares the numerical results obtained using the full lab-frame Hamiltonian (Eq.~\!\eqref{eq:Hamiltonian_Full}) and the rotating-frame approximation (Eq.~\!\eqref{eq:eq003}). We observe that even under a relatively strong drive amplitude of 100~MHz, simulations of the full Hamiltonian exhibit only minor high-frequency oscillations in state populations, with maximum deviations below 0.5\% compared to the RWA results. Critically, the obtained gate fidelity differences remain below $10^{-4}$, demonstrating negligible operational impact from counter-rotating terms. These results validate RWA-based simulations while confirming gate robustness in our parameter regime.
\color{black}

\section{Suppression of Static ZZ Coupling via Continuous-Wave Stark Drives}
\label{app:CWStark}

\begin{figure*}[!htbp]
	\centering
	{\includegraphics[width = 0.85\linewidth]{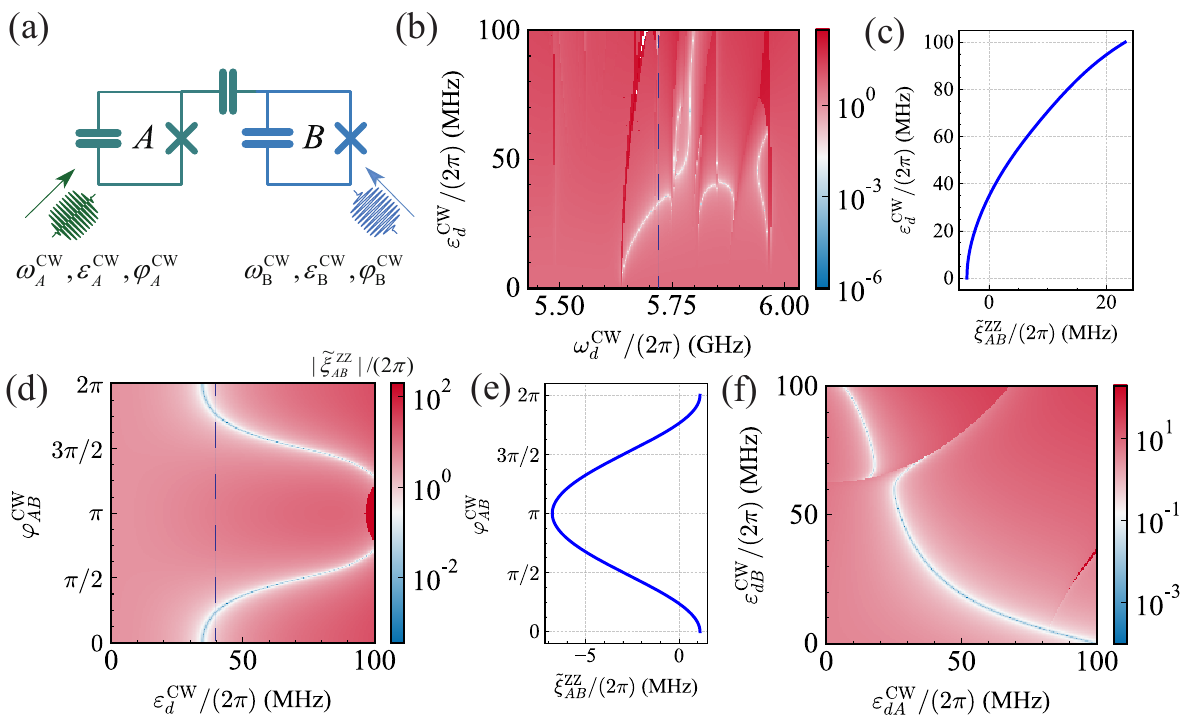}}
	\caption{Modulation of ZZ interaction via continuous-wave (CW) Stark drives. (a) Circuit schematic of coupled transmons ($q_{\sss A}$, $q_{\sss B}$) with simultaneous CW drives. Parameters: $\omega_{\mathrm{\sss A,B}}/2\pi = \{5.641, 6.517\}$ GHz, $\alpha_{\mathrm{\sss A,B}}/2\pi = \{-300, -381\}$ MHz, $g_{\mathrm{\sss AB}}/2\pi = 40$ MHz. 
	(b) $\tilde{\xi}_{\mathrm{\sss A,B}}^{\sss ZZ}$ as a function of drive   frequency $\omega_d^{\mathrm{\sss CW}}$ ($\omega_d^{\mathrm{\sss CW}}=\omega_{\sss dA}^{\mathrm{\sss CW}}=\omega_{\sss dB}^{\mathrm{\sss CW}}$)
	and amplitude $\varepsilon^{\mathrm{\sss CW}}_d$
	($\varphi^{\mathrm{\sss CW}}_{\mathrm{AB}} = 0$). Full suppression occurs in regions where $\omega_d^{\mathrm{\sss CW}} > \omega_{\mathrm{\sss B}}$. 
	(c) Line trace of (b) at fixed $\omega_d^{\mathrm{\sss CW}}/2\pi = 5.729$ GHz ($\Delta^{\mathrm{\sss CW}}_{\mathrm{\sss dA}}/2\pi = 89$ MHz), showing transition from static coupling ($\sim\!-3.75$ MHz) to full suppression.
	(d) $\tilde{\xi}_{ZZ}$ versus relative phase $\varphi^{\mathrm{\sss CW}}_{\mathrm{AB}}$ and drive amplitude at fixed $\omega_d^{\mathrm{\sss CW}}/2\pi = 5.729$ GHz, highlighting distinct phase dependence for $\varepsilon^{\mathrm{\sss CW}}_d$ $>$ 40 MHz.  
	(e) Sinusoidal phase modulation at $\varepsilon^{\mathrm{\sss CW}}_d = 40$ MHz.
	(f) Independent amplitude control of $\varepsilon^{\mathrm{\sss CW}}_{\sss dA}$ and $\varepsilon^{\mathrm{\sss CW}}_{\sss dB}$ for ZZ interaction cancellation.
	}
	\label{fig:CWstark}
\end{figure*}
\begin{figure}[!htbp]
    \centering
	\includegraphics[width=0.48\textwidth]{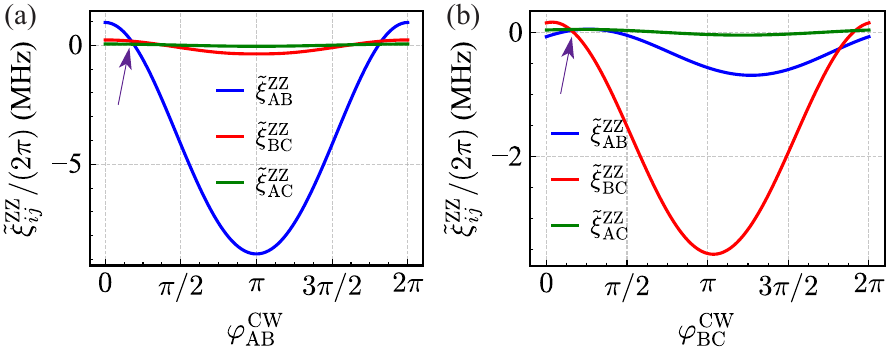}
	\caption{Static ZZ modulation in a three-transmon system via CW drives. (a) Residual ZZ coupling $\tilde{\xi}^{\mathrm{\sss ZZ}}_{ij}$ versus relative phase $\varphi_{\mathrm{\sss AB}}^{\mathrm{\sss CW}}$ at fixed drive amplitudes $\varepsilon_{\mathrm{\sss A,B,C}}^{\mathrm{\sss CW}}/2\pi = \{35, 55, 40\}$ MHz and frequency $\omega_d^{\mathrm{\sss CW}}/2\pi = 5.729$ GHz. The purple arrow indicates complete suppression of $\tilde{\xi}_{\mathrm{\sss AB}}^{\mathrm{\sss ZZ}}$. (b) Simultaneous minimization of NN ($\tilde{\xi}_{\mathrm{\sss AB}}^{\mathrm{\sss ZZ}}$, $\tilde{\xi}_{\mathrm{\sss BC}}^{\mathrm{\sss ZZ}}$) and NNN ($\tilde{\xi}_{\mathrm{\sss AC}}^{\mathrm{\sss ZZ}}$) couplings achieved through coordinated phase adjustment of $\varphi_{\mathrm{\sss BC}}^{\mathrm{\sss CW}}$, with optimal suppression point marked by the purple arrow.}
    \label{fig:CWstarkABC}
\end{figure}

As discussed in Section~\ref{sec:transition}, when fixed-frequency transmon qubits are operated in the large-detuning regime ($|\Delta_{\sss AB,BC}| \gg g_{\sss AB,BC}$) to effectively suppress static NNN ZZ couplings, the static NN ZZ coupling persists at significant MHz-scale magnitudes. These residual ZZ couplings (Eq.~\eqref{eq:eqZZNNij}) substantially degrade single-qubit gate fidelity by inducing state-dependent frequency shifts that compromise simultaneous operations. Thus, high-fidelity gates are typically complicated by the requirement to engineer cancellation or suppression of this ZZ interaction.  While strategies of ZZ suppression exist, such as coupling reduction (compromising gate speed), tunable couplers~\cite{PhysRevLett.127.200502}, opposite-anharmonicity pairs~\cite{PhysRevLett.129.060501}, or multipath couplers~\cite{PhysRevA.102.062408}, they generally require additional tunable components incompatible with fixed-frequency architectures.

To address this limitation, we adopt an established  microwave-based suppression method employing simultaneous continuous-wave (CW) Stark drives to cancel static ZZ interactions~\cite{PhysRevLett.127.200502,PhysRevLett.129.060501,PhysRevA.102.062408,Goss2022}. When applied off-resonance to coupled qubits, these drives generate conditional Stark shifts that suppress ZZ interactions by orders of magnitude~\cite{PhysRevLett.127.200502,PhysRevLett.129.060501}. While previous implementations have exclusively operated in the straddling regime ($\Delta < \alpha$), we extend this approach to the large-detuning regime relevant to our system. For two coupled qubits subject to simultaneous CW drives at frequency $\omega_d^{\sss \mathrm{CW}}$, the driven Hamiltonian after moving into the frame of drive (with fast-rotating terms neglected) is expressed  as,
\begin{equation}\label{eq:CWRWAHamiltonian}
\begin{aligned}
    H =& \sum_{i=\mathrm{\sss A,B}}  \Delta^{\mathrm{\sss CW}}_{d,i} a_i^\dagger a_i + \frac{\eta_i}{2} a_i^\dagger a_i^\dagger a_i a_i  + g_{\mathrm{\sss A,B}} \left( a_{\sss A}^\dagger a_{\sss B} + a_{\sss A} a_{\sss B}^\dagger \right) \\ 
    &+\varepsilon^{\mathrm{\sss CW}}_{d,i} \left( e^{i\varphi^{\sss \mathrm{CW}}_i} a_i + e^{-i\varphi^{\sss \mathrm{CW}}_i} a_i^\dagger \right),
\end{aligned}
\end{equation}
where $\varepsilon^{\mathrm{\sss CW}}_{d,i}$ and $\varphi^{\sss \mathrm{CW}}_{i}$ denote the drive amplitude and phase for qubit $i$, and $\Delta^{\mathrm{\sss CW}}_{d,i} = \omega_i - \omega^{\sss \mathrm{CW}}_d$ defines  the drive-qubit detuning. Perturbative analysis yields the modulated ZZ strength~\cite{PhysRevLett.127.200502,PhysRevLett.129.060501},
\begin{equation}\label{eq:CWRWAAnalytic}
\tilde{\xi}_{\sss \mathrm{AB}}^{\sss ZZ} = \xi_{\sss \mathrm{AB}}^{\sss ZZ,0} + \frac{8 g_{\mathrm{\sss AB}} \alpha_{\mathrm{\sss A}} \alpha_{\mathrm{\sss B}} \varepsilon^{\sss \mathrm{CW}}_{{\sss dA}} \varepsilon^{\mathrm{\sss CW}}_{{\sss dB}} \cos\left( \varphi^{\sss \mathrm{CW}}_{\mathrm{\sss AB}} \right)}{\Delta^{\mathrm{\sss CW}}_{\sss dA} \Delta^{\mathrm{\sss CW}}_{\sss dB} \left( \Delta^{\mathrm{\sss CW}}_{\sss dA} + \alpha_{\mathrm{\sss A}} \right) \left( \Delta^{\mathrm{\sss CW}}_{\sss dB} + \alpha_{\mathrm{\sss B}} \right)},
\end{equation}
where $\varphi^{\sss \mathrm{CW}}_{\mathrm{\sss AB}} \equiv \varphi^{\sss \mathrm{CW}}_{\mathrm{\sss A}} - \varphi^{\sss \mathrm{CW}}_{\mathrm{\sss B}}$ is the relative drive phase. The first term $\xi_{\mathrm{\sss AB}}^{ZZ,0}$ represents the static ZZ coupling, while the second term introduces a tunable component that depends on drive amplitudes ($\varepsilon_{d,i}^{\mathrm{\sss CW}}$), detunings ($\Delta_{{d,i}}^{\mathrm{\sss CW}}$), and relative phase ($\varphi_{\mathrm{\sss AB}}^{\mathrm{\sss CW}}$). This provides independent control dimensions for targeted ZZ suppression or modulation.

To validate the suppression approach, we performed numerical simulations on an isolated transmon pair ($q_{\sss A}$, $q_{\sss B}$) using parameters consistent with the main text: $\omega_{\mathrm{\sss A,B}}/2\pi = \{5.641, 6.517\}$ GHz, $\alpha_{\mathrm{\sss A,B}}/2\pi = \{-300, -381\}$ MHz, and $g_{\mathrm{\sss AB}}/2\pi = 40$ MHz. These parameters generate a static NN ZZ coupling of $\xi_{\mathrm{\sss AB}}^{ZZ,0}/2\pi \approx -3.7$ MHz. 
Full diagonalization of Eq.~\eqref{eq:CWRWAHamiltonian} reveals a multifaceted landscape (Fig.~\ref{fig:CWstark}) of $\tilde{\xi}_{\sss \mathrm{AB}}^{\sss ZZ}$. Each transmon is modeled with 7 energy levels in the diagonalization.  Fig.~\ref{fig:CWstark}(b) demonstrates that complete cancellation of $\tilde{\xi}_{\mathrm{\sss AB}}^{\mathrm{\sss ZZ}}$ occurs primarily when $\omega_{\sss d}^{\mathrm{\sss CW}} > \omega_{\mathrm{\sss B}}$  and sizeable modulation modulation can be achieved across a broad range of frequencies. 
The sharp resonant features in this regime indicate interactions with higher transmon energy levels. 
 Consistent with Eq.~\eqref{eq:CWRWAAnalytic}, reduced drive detunings $\Delta^{\mathrm{\sss CW}}_{\mathrm{\sss A}}$ lower the required cancellation amplitude, though excessively small values risk resonant interference with gate operations and increased leakage. We therefore implement a conservative $\Delta^{\mathrm{\sss CW}}_{{\sss dA}}/2\pi = 89$ MHz ($\omega_{\sss d}^{\mathrm{\sss CW}}/2\pi/2\pi = 5.729$ GHz), achieving complete suppression at $\varepsilon^{\mathrm{\sss CW}}_{\sss d} \approx 40$ MHz of $\tilde{\xi}_{\sss \mathrm{AB}}^{\sss ZZ}$ (Fig.~\ref{fig:CWstark}(c)). 
 The relative drive phase $\varphi^{\mathrm{\sss CW}}_{\mathrm{AB}}$ serves as a critical control knob. As illustrated  in Fig.~\ref{fig:CWstark}(d), a sinusoidal phase dependence (Fig.~\ref{fig:CWstark}(e)) arises for drive amplitudes $\varepsilon^{\mathrm{\sss CW}}_{\sss d}/2\pi > 37$ MHz, showcasing its effectiveness as a primary tuning parameter for ZZ crosstalk suppression. This phase control also enables versatile reactivation of the ZZ interaction when required, providing essential flexibility for implementing CPhase and CZ gate operations. Figure~\ref{fig:CWstark}(e) reveals the predicted $\cos\varphi^{\mathrm{\sss CW}}_{\mathrm{\sss AB}}$ dependence at $\varepsilon^{\mathrm{\sss CW}}_{\sss d}/2\pi = 40$ MHz. 
 Figure~\ref{fig:CWstark}(f) plots the dependence of $\tilde{\xi}_{\sss \mathrm{AB}}^{\sss ZZ}$ on $\varepsilon^{\mathrm{\sss CW}}_{{\sss dA}}$ and $\varepsilon^{\mathrm{\sss CW}}_{{\sss dB}}$, clearly showing the $\tilde{\xi}_{\sss \mathrm{AB}}^{\sss ZZ} \propto \varepsilon^{\mathrm{\sss CW}}_{{\sss dA}} \varepsilon^{\mathrm{\sss CW}}_{{\sss dB}}$ relationship anticipated from Eq.~\eqref{eq:CWRWAAnalytic}. This affords the flexibility to optimize drive parameters under experimental constraints while maintaining cancellation conditions.

Building upon the two-qubit validation, we extend our numerical simulations to a three-transmon system using parameters consistent with the main text. With the drive frequency fixed at $\omega_{\sss d}^{\mathrm{\sss CW}}/2\pi = 5.729$ GHz (corresponding to detunings $\Delta^{\mathrm{\sss CW}}_{\mathrm{\sss A}}/2\pi = 89$ MHz and $\Delta^{\mathrm{\sss CW}}_{\mathrm{\sss B}}/2\pi = 221.87$ MHz), and drive amplitudes $\varepsilon^{\mathrm{\sss CW}}_{\mathrm{\sss A,B,C}}/2\pi = \{35, 55, 40\}$ MHz, we systematically investigate the residual ZZ couplings $\tilde{\xi}_{\mathrm{\sss AB}}^{\mathrm{\sss ZZ}}$, $\tilde{\xi}_{\mathrm{\sss BC}}^{\mathrm{\sss ZZ}}$, and $\tilde{\xi}_{\mathrm{\sss AC}}^{\mathrm{\sss ZZ}}$. As shown in Figure~\ref{fig:CWstarkABC}, optimal suppression of both NN and NNN ZZ interactions is achieved through coordinated adjustment of the relative microwave drive phases. The results demonstrate that CW Stark driving represents a potent technique for the effective suppression of  static ZZ crosstalk in fixed-coupling, large-detuning architectural designs. By leveraging microwave drive phase and amplitude freedom, this technique enables  comprehensive NN and NNN ZZ cancellation without hardware modifications. The framework dynamically reconfigures to suppress ZZ for single-qubit operations or activate it for controlled-phase gates, offering significant scalability advantages. 
In parallel, Huang et al.~\cite{PhysRevApplied.22.034007} recently demonstrated a resonator-mediated strategy for large-detuning architectures where a passive resonator detuned by $\sim$5 GHz from the qubits simultaneously suppresses static ZZ coupling (approximately 6 MHz) and enables ultrafast entangling gates. This approach leverages resonator-mediated interactions to preserve qubit coherence in fixed-frequency systems, achieving 40-ns CNOT and 140-ns adiabatic CZ gates with simulated fidelities exceeding 99.99\%. The introduction of this fixed coupling element not only provides an effective solution for fixed-qubit architectures but also maintains compatibility with our three-qubit gate implementations, offering a comprehensive approach for direct high-fidelity gate operations.

\color{black}

\section{Details of the Evolution Operator Simulation }
\label{app:EvolutionOperator}
In the non-interacting regime $g_{ij} = 0$, the system governed by Eq.~\!\eqref{eq:eq001} exhibits bare eigenstates $|ijk\rangle$, where $i, j, k \in \mathbb{N}$ label the individual transmon levels. When couplings are introduced, the eigenstates of $H_0$ evolve into dressed states $|\widetilde{ijk}\rangle$, characterized by slight energy shifts relative to their bare counterparts and a substantial overlap with the bare state. Each dressed eigenstate is labeled as $|\widetilde{ijk}\rangle$ according to the bare state with which it shares maximal overlap. The computational subspace is spanned by states $|\widetilde{ijk}\rangle$ with $i, j, k \in \{0, 1\}$.

To numerically construct the evolution operator, we simulate the time evolution of the system under Eq.~\!\eqref{eq:eq003} using QuTiP~\!\cite{JOHANSSON20121760}, initializing the dynamics with the computational states $|\widetilde{\psi_i}\rangle$. The evolved states $|\widetilde{\psi^\prime_i}\rangle$ are projected onto the basis states to construct the operation matrix $\widetilde{U}$, whose elements are given by $U_{ij} = \langle \psi_i | \psi^\prime_j \rangle$.  Note that $\widetilde{U}$ is not a unitary matrix in general due to leakage erros~\!\cite{PhysRevApplied.21.044035,PhysRevB.96.024504,PhysRevApplied.18.034038,PhysRevA.108.032615}. From this procedure, we obtain the microwave-activated operation $\widetilde{U}$ and then implement a phase optimization protocol to derive the phase compensated unitary $U^\prime = U_{\text{phase}} \widetilde{U}(\phi_{d})$. The performance of the implemented gate is quantified using the averaged gate fidelity metric~\!\cite{Phys.Lett.A367.47},
\begin{equation}
F = \frac{\left|\mathrm{Tr}(U_{\mathrm{ideal}}^{\dagger} U^\prime)\right|^{2} + \mathrm{Tr}\left(U^{\prime \dagger} U^\prime\right)}{d(d + 1)},
\end{equation}
where $d = 8$ represents the Hilbert space dimension of the three-qubit system, and $U_{\mathrm{ideal}}$ corresponds to the ideal gate defined in Eq.~\!\eqref{eq:eq006}.  The phase-compensated unitary $U^\prime$ is constructed as $U^\prime = U_{\text{phase}} \widetilde{U}(\phi_{d})$, with $\phi_{d}$ denotes the pulse phase. Here, $U_{\text{phase}}$ comprises three single-qubit virtual Z gates and two CPhases for $AB$ and $BC$.


\begin{figure}[!htbp]
	\centering 
	\includegraphics[width=0.48\textwidth]{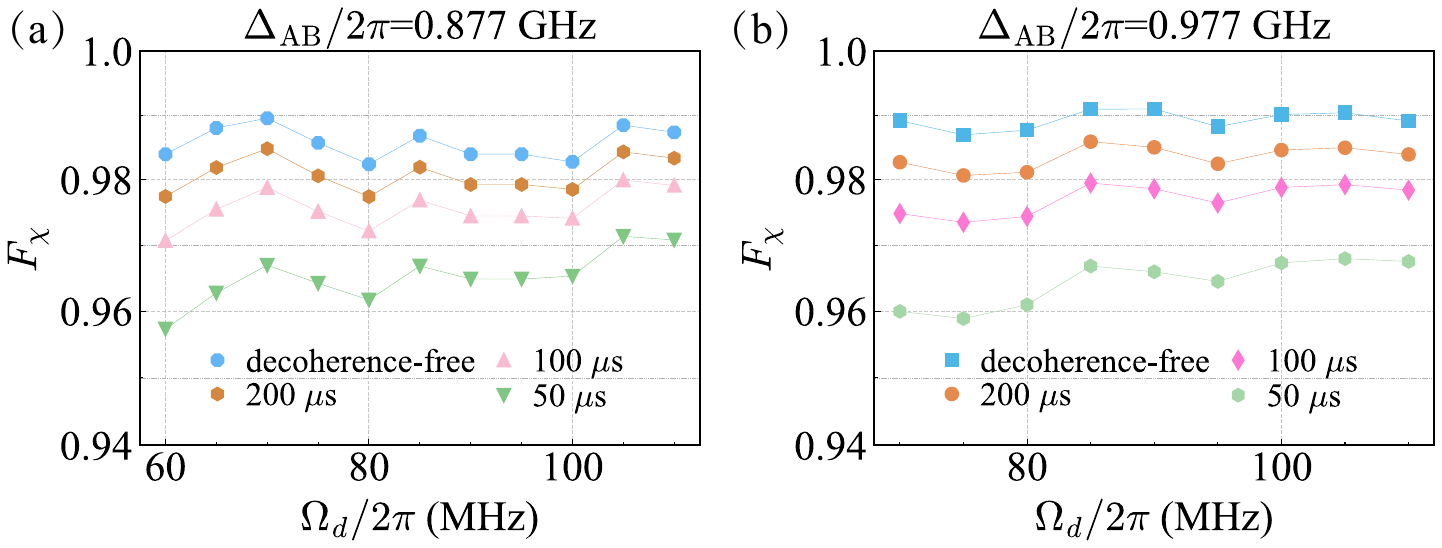}
	\caption{Gate performance under decoherence conditions for large detunings: $\Delta_{\mathrm{AB}} = 0.877$ GHz (a) and $\Delta_{\mathrm{AB}} = 0.977$ GHz (b). Process fidelity exceeds 98.5\% for $\Delta_{\mathrm{AB}} = 0.877$ GHz and reaches $>$99.0\% at $\Delta_{\mathrm{AB}} = 0.977$ GHz under decoherence-free conditions. Under realistic decoherence ($T_1 = T_2^* =$ 200, 100, 50 $\mu$s), fidelity decreases but remains near 96\% even at 50 $\mu$s.
	}
	\label{fig:Figs_qpt}
\end{figure}

\section{Details of Process Fidelity Simulations}
\label{app:ProcessFidelity}

We numerically simulate the system's time evolution by solving the Lindblad master equation (Eq.~\!\eqref{eq:eq003}) in QuTiP~\!\cite{JOHANSSON20121760},
\begin{equation}
\dot{\hat{\rho}}(t) = -i[\hat{H}_{\mathrm{RF}}(t), \hat{\rho}] + \sum_{\alpha} \left( \hat{L}_{\alpha} \hat{\rho} \hat{L}_{\alpha}^{\dagger} - \frac{1}{2} \left\{ \hat{L}_{\alpha}^{\dagger} \hat{L}_{\alpha}, \hat{\rho} \right\} \right),
\end{equation}
where $\hat{\rho}$ is the density matrix, $\hat{H}_{\mathrm{RF}}$ represents the system Hamiltonian, and the Lindblad operators $\hat{L}_1 = \sqrt{\Gamma_1} |0\rangle \langle 1|$ (relaxation) and $\hat{L}_\phi = \sqrt{\Gamma_\phi / 2} (|0\rangle \langle 0| - |1\rangle \langle 1|)$ (pure dephasing) model decoherence. As our gate protocol operates exclusively within the computational subspace, decoherence effects are restricted to transitions between $|0\rangle$ and $|1\rangle$ states.  

To reconstruct the process matrix for the microwave-activated three-qubit gate, we prepare 64 initial states spanning the three-qubit Hilbert space: $\{|0\rangle, |1\rangle, (|0\rangle + |1\rangle)/\sqrt{2}, (|0\rangle - i|1\rangle)/\sqrt{2}\}^{\otimes 3}$. 
Each initial state evolves under the driven Hamiltonian $\hat{H}_{\mathrm{RF}}$ with the microwave control field, where the pulse phase $\phi$ remains fixed at values pre-optimized through average fidelity maximization in Appendix~\ref{app:EvolutionOperator}. Following evolution, the resulting density matrices $\rho_{\text{final}}$ undergo unitary correction through the $U_{\mathrm{phase}}$ operation:
\begin{equation}
\begin{aligned}
   U_{\mathrm{phase}} &= \left(R^z(\varphi_{\sss A}) \otimes R^z(\varphi_{\sss B}) \otimes R^z(\varphi_{\sss C})\right) \cdot \\
   & \left(\mathrm{CPhase}(\theta_{\sss AB}) \otimes \mathrm{CPhase}(\theta_{\sss BC})\right),
\end{aligned}
\end{equation}
which mitigates phase errors using five adjustable parameters $\{\varphi_{\sss A}, \varphi_{\sss B}, \varphi_{\sss C}, \theta_{\sss AB}, \theta_{\sss BC}\}$. These parameters are initialized using values obtained from average fidelity optimization in Appendix~\ref{app:EvolutionOperator}. The quantum process matrix $\chi$ is reconstructed via maximum likelihood estimation using the relation, $\rho_{\text{final}} = \sum_{n,m} \chi_{nm} E_n \rho_{\text{ini}} E_m^\dagger$, where the basis operators $\{E_n\}$ comprise the Pauli tensor product set $\{I, \sigma_x, -i\sigma_y, \sigma_z\}^{\otimes 3}$~\cite{Chen2022}. The process fidelity is then computed as $F_\chi = \text{Tr}(\chi_{\text{ideal}}\chi)$~\!\cite{PhysRevX.11.021058,PRXQuantum.3.037001}, with $\chi_{\text{ideal}}$ representing the ideal process matrix for the target gate operation $U_{\mathrm{ideal}}$~\cite{PhysRevX.11.021058}. To enhance fidelity, we implement a targeted optimization protocol where the five $U_{\mathrm{phase}}$ parameters undergo gradient-based refinement to maximize $F_\chi$, while all operational parameters (including $\phi_{d}$) remain fixed. The optimized $F_\chi$ values across decoherence regimes are reported in Fig.~\ref{fig:Figs_qpt}.

For large detuning $\Delta_{\mathrm{\sss AB}} = 0.877$ GHz, the process fidelity exceeds 98.5\%, while increasing the detuning to $\Delta_{\mathrm{\sss AB}} = 0.977$ GHz further elevates fidelity beyond $99.0\%$ under decoherence-free conditions.
When incorporating realistic decoherence rates with relaxation and dephasing times $T_1 = T_2^* = 200$, 100, and 50 $\mu$s (shown in Fig.~\ref{fig:Figs_qpt}), the fidelity gradually decreases. Specifically, it remains near 98\% at a dephasing time of $T_2^*$=200 $\mu$s, exceeds $>$97\% at 100 $\mu$s, and stays near 96\% even at 50 $\mu$s. 
Given recent advances in fabrication techniques, fixed-frequency transmon qubits, which are among the most coherent superconducting elements, now routinely exceed 200 $\mu$s coherence times~\cite{Place2021.0.3ms,Wang2022.0.5ms,Bal2024,interfacialdielectricloss2024,Bu2025,Tuokkola2025T11ms}. The achieved fidelity thus demonstrates robust performance against relaxation and dephasing errors, validating our three-qubit gate protocol's compatibility with realistic superconducting platforms.


%

\end{document}